%% file: main.tex
\documentclass[preprint,12pt]{elsarticle}
\usepackage{amsmath}
\usepackage{graphicx,psfrag,epsf}
\usepackage{multirow}
\usepackage{graphics}
\usepackage{enumerate}
\usepackage{natbib}
\usepackage{mathrsfs}
\usepackage{hyperref}
\usepackage{autobreak}
\usepackage{amsfonts,amssymb}
\usepackage{url} 

\DeclareMathOperator*{\argmin}{arg\,min}

\usepackage{amsthm}

\newtheorem{theorem}{Theorem}
\newtheorem{assumption}{Assumption}
\newtheorem{lemma}{Lemma}
\newtheorem{corollary}{Corollary}

\newcommand{\bbeta}{\mbox{\boldmath $\beta$}}

\newcommand{\btheta}{\mbox{\boldmath $\theta$}}
\newcommand{\bLambda}{\mbox{\boldmath $\Lambda$}}

\def\bfB{{\bf B}}

\def\bfC{{\bf C}}

\def\bfH{{\bf H}}

\def\bfI{{\bf I}}

\def\bfX{{\bf X}}

\def\bfz{{\bf z}}

\def\bbG{{\mathbb G}}
\def\bbV{{\mathbb V}}
\def\bbE{{\mathbb E}} 

\journal{Spatial Statistics}

\begin{document}
\begin{frontmatter}

\title{Fused Spatial Point Process Intensity Estimation with Varying Coefficients on Complex Constrained Domains}

\author[inst1,inst2]{Lihao Yin}
\affiliation[inst1]{organization={Department of Statistics, Texas A\&M University},  
            city={College Station}, 
            state={TX},
            country={USA}}
\author[inst1]{Huiyan Sang} 

\affiliation[inst2]{organization={Institute of Statistics and Big Data, Renmin University},  
            city={Beijing}, 
            state={},
            country={China}}

\begin{abstract}
The availability of large spatial data geocoded at accurate locations has fueled a growing interest in spatial modeling and analysis of point processes.  The proposed research is motivated by the intensity estimation problem for large spatial point patterns on complex domains in $\mathbb{R}^2$ (e.g., domains with irregular boundaries, sharp concavities, and/or interior holes due to geographic constraints) and linear networks, where many existing spatial point process models suffer from the problems of ``leakage" and computation. We propose an efficient intensity estimation algorithm to estimate the spatially varying intensity function and to study the varying relationship between intensity and explanatory variables on complex domains. The method is built upon a graph regularization technique and hence can be flexibly applied to point patterns on complex domains such as regions with irregular boundaries and holes, or linear networks. An efficient proximal gradient optimization algorithm is proposed to handle large spatial point patterns.  We also derive the asymptotic error bound for the proposed estimator. Numerical studies are conducted to illustrate the performance of the method.  Finally,  we apply the method to study and visualize the intensity patterns of the accidents on the Western Australia road network,  and the spatial variations in the effects of income, lights condition, and population density on the Toronto homicides occurrences. 
\end{abstract}

\begin{keyword}  Graph regularization, Intensity estimation, Line network, Spatial point pattern, Varying coefficient models.  
\end{keyword} 

\end{frontmatter} 
 
\section{Introduction} 
\label{sec:intro}

Numerous problems in geosciences, social sciences, ecology, and urban planning nowadays involve extensive amounts of spatial point pattern data recording event occurrence. Examples include locations of invasive species, pick-up locations of Taxi trips, addresses of 911 calls, and traffic accidents on roads, to name a few. In many such applications, the primary problem of interest is to characterize the probability of event occurrence. In the presence of additional covariates information, another problem of interest is to study the effect of these covarites on event occurrence probability, considering the spatial dependence of observations. Spatial point process models have been widely used for the analysis of point patterns, in which the intensity function, denoted as $\rho(u)$, is used to describe the likelihood for an event to occur at location $u$.  

\begin{figure}
\begin{center}
\includegraphics[width=6in]{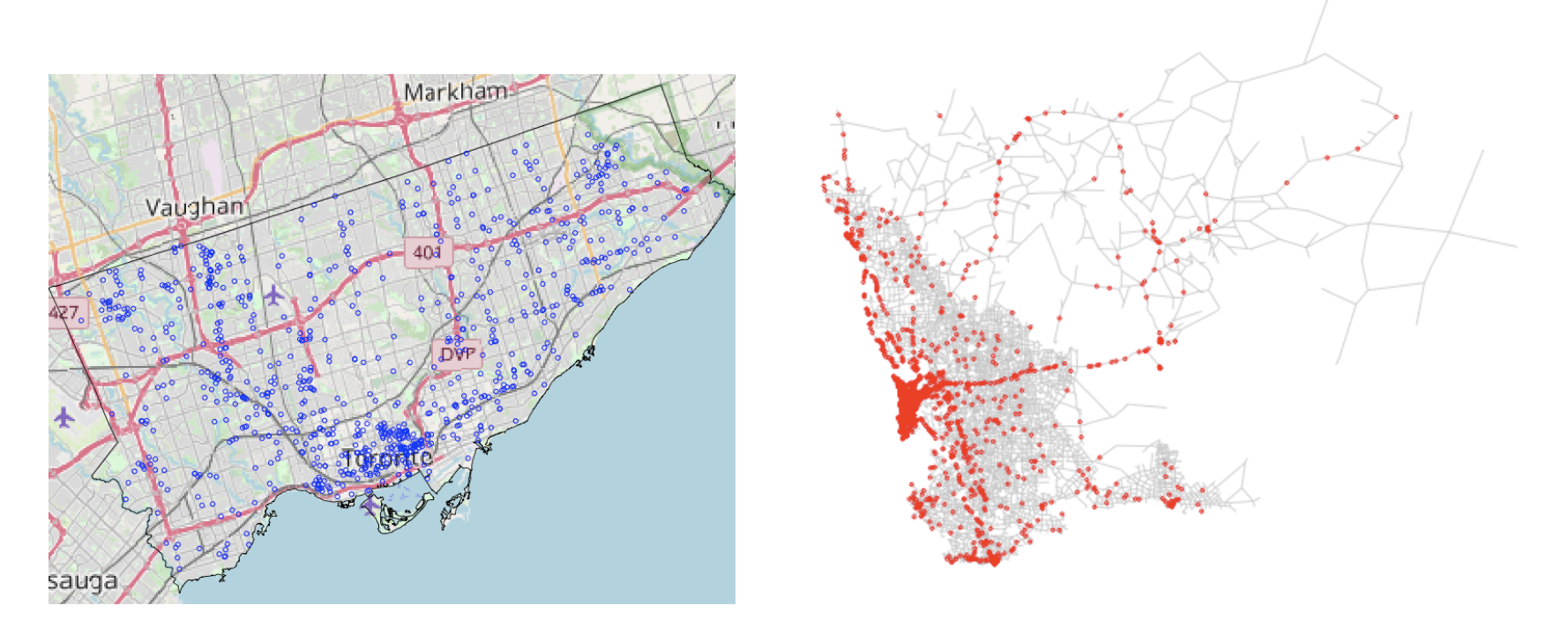}
\end{center}
\caption{Left: Map of homicide locations in Toronto during $2000-2014$; Right: Traffic networks and road accidents on Western Australia in 2011;}
\label{fig1}
\end{figure}

In practice, many spatial point patterns data are collected over complex domains with irregular boundaries, peninsulas, interior holes, or network geographical structures. In this paper, we consider two motivating data examples on such complex domains. The first one is the traffic accident locations on the Western Australia road network shown in the right panel of Figure \ref{fig1}, where the interest lies in studying the spatial variation of accident occurrences. The left panel in Figure \ref{fig1} shows the homicides that occurred in Toronto, where the city boundary has a very irregular shape especially near Toronto islands. The Toronto data set also includes several additional covariates such as the records of average income, night lights, and population density. Therefore, the questions of interest include not only the intensity of crime events but also the relationships between crime intensity and regional characteristics. In particular, for a large city like Toronto, we may expect that such relationships can vary, and in some places rather abruptly, across the study domain.

Thus far, many methods have been introduced to model the first-order spatially varying intensity function $\rho(u)$. Popular point process models include the spatial Poisson point processes, the log-Gaussian Cox Processes, and the Gibbs point processes. See a review by \cite{moller2007modern}. The intensity estimations of these models are often done using maximum composite likelihoods~\citep{guan2006composite}, estimating equations~\citep{guan2015quasi} or Bayesian inference methods~\citep{leininger2017bayesian,gonccalves2018exact,shirota2019scalable}.  
Nonparametric methods have also been widely used for estimating the spatially varying intensity functions, including the edge-corrected kernel smoothing estimators by \cite{diggle1985kernel,jones1996brief}, the Voronoi estimator by \cite{barr2010voronoi}  using the inverse of the area of the Voronoi cell for each observed location, and a local likelihood estimation procedure in analogy to geographically weighted regression by \cite{fotheringham2003geographically}.

Yet, statistical analysis of point patterns on complex domains presents severe challenges to many of the classical point process models reviewed above. Mainly, the commonly adopted Euclidean assumption underpinning some of these methods no longer holds for point patterns on complex domains.  For example, two locations on a road network that are close by Euclidean distance may actually lie on two separate roads.  Moreover, the large data size will aggravate the challenges in modeling point patterns on complex domains. There is a great need to develop spatial point pattern analysis tools that are  computationally efficient 
to solve the so called ``leakage" problem encountered on complex domains. 

For point patterns that occur along a line network, a particular type of complex domain,  a number of methods have been developed recently. Kernel estimators of the intensity function on a line network were investigated in \cite{mcswiggan2017kernel,moradi2018kernel,rakshit2019fast}, adapting the idea of edge-correction using path lengths. Other variations of kernel density estimation methods are reviewed in \cite{baddeley2020analysing}. It is known that kernel estimators are, by nature, more suitable for estimating relatively smooth intensity functions because of the use of smoothing kernel functions. When intensity function exhibits discontinuities and abrupt changes in space, as discussed in \cite{baddeley2020analysing},  piece-wise constant estimators become an appealing alternative as they have a strong adaptivity to changes.  One research in this direction is the aforementioned Voronoi estimator by \cite{barr2010voronoi}.  However, the method suffers from the high variance in the estimator. To reduce the variance, \cite{moradi2019resample} extended it by a bootstrap resample smoothing procedure. Recently, \cite{bassett2019fused} proposed to estimate the density of points on a network as opposed to the intensity function based on a total variation regularization method. While each represents advancements in estimating intensity or density of points, none has incorporated spatial covariates in estimation.

When spatial covariates are available, various methods \citep{baddeley2012nonparametric,McSwiggan2019} have been developed to incorporate covariate information with the goal to investigate the effect of spatial covariates on point patterns. However, to the best of our knowledge, there has been very limited work for dealing with varying regression coefficients for spatial point patterns, even in the simpler case where point patterns are observed in the Euclidean space.  One notable exception is the work by \cite{pinto2015point}, which modeled the regression coefficients as a multivariate Gaussian process in a similar fashion as the spatially varying coefficients (SVC) linear regression model proposed by \cite{gelfand2003spatial}.  Despite the model richness and flexibility, the SVC model is known to involve heavy computation in the presence of large spatial data due to the requirement of Metropolis MCMC and the need to invert a large covariance matrix. The intractability of the likelihood function of the spatial Poisson process further aggravates the issue. To address the computation issue, \cite{pinto2015point} partitioned the study region into a small number of subregions according to administrative areas and assumed that latent spatial random effects take constant values within each subregion.  However, in some applications, such a pre-determined partition may be unavailable or fail to accurately reflect the complex underlying environmental and geological conditions.  

In light of these limitations in the current literature, we develop a simple yet effective approach based on a fused lasso regularization method on a graph for the estimation of piece-wise constant spatial intensity functions. We propose penalties on regression coefficients to encourage sparsity on the differences among regression coefficients that are close in space. The fused lasso methods have gained increasing popularity owning to its flexibility of learning clustered structures. However, to our knowledge, there is limited work that has investigated its performance for point pattern data analysis. 
In addition, we extend the approach to a piece-wise constant coefficient spatial point process model when explanatory variables are available, to model the varying relationships between point patterns and covariates.  We formulate the estimation problem into penalized Poisson-based and Logistic based composite likelihoods optimizations, for which we solve by an efficient proximal gradient algorithm. We tailor the algorithm to utilize spatial graph structures such that the method is applicable for dealing with large spatial point patterns. The choices of graphs play important roles in the modeling and computation of fused lasso problems. We consider various spatial graphs to represent spatial geometry of complex domains and compare their performance. We also make a theoretical contribution by studying the asymptotic properties of the proposed estimator. Finally, we introduce this method to the analysis of the Western Australia accident data and the Toronto homicides data. The results of our analysis reveal several interesting clustering patterns of traffic accidents and the spatial crime distribution in relation to a number of key environmental, social, and economic variables. The R code is included as a supplementary file.  

The paper is organized as follows. In Section~\ref{sec:background}, we review the basic mathematical formulations and definitions of spatial point processes. We then introduce our method in Section \ref{sec:SVCI}, followed by the computation algorithm in \ref{sec:Computation} and theoretical results in \ref{sec:theory}. 
Sections \ref{sec:simulation} and \ref{sec:realdata} include the simulations to illustrate the model performance and the applications to the two real data sets. We offer conclusions and discussions in Section \ref{sec:conclusion}. The proof of Theorem 1 and additional implementation details and numerical results are in Appendix.  

\section{Preliminaries}\label{sec:background} 
\subsection{Observation Domain}
In this study, we consider spatial points on two important types of observation domains. 
The first type is a bounded domain $D \subset \mathbb{R}^2$ that can be fully covered by finitely many rectangles. The commonly assumed planar window  $[a_1,a_2]\times [b_1,b_2]$ is a special case of this type. For any locations $u_1,u_2$ in a planar window, the Euclidean distance is used to measure the distance between two locations, denoted as $d(u_1,u_2)$. One example of this type is given in Figure \ref{fig1}, where the observation domain is the city of Toronto, which has irregular city limit boundaries. For any Borel subset $B \subseteq D$, the Lebesgue measure $|B|$ is the area of $B$.  

In the second type, we assume $D$ is a linear network. Let $[u,v]=\{tu+(1-t)v:0\leq t\leq 1\}$ denotes a line segment in the plane with endpoints $u,v\in\mathbb{R}^2$. A linear network is defined as the finite union $D=\cup_{i=1}^\eta[u_i,v_i]$ of line segments $[u_1,v_1],\cdots,[u_\eta,v_\eta]$ embedded in the same plane. One commonly used distance $d(u_1,u_2)$ is the shortest-path distance between $u_1$ and $u_2$ on the network. For any subset $B\subseteq D$, the measure $|B|$ represents the total length of all segments in $B$. An example of the line network is shown in the right panel of Figure \ref{fig1}, where the road network in the state of Western Australia is drawn in grey lines, and red points mark the traffic accident locations in 2011.

\subsection{Spatial Point Processes}
Let $\mathbf{X}$ be a spatial point process on $D$ with the locally finite property, i.e., the random cardinality $N_{\mathbf{X}}(B)=\#\{u:u\in \mathbf{X}\cap B\}$ is almost surely finite for any $B\subset D$. 
Assume that, for any bounded $B\subset D$, if there exits a non-negative and locally integrable function $\rho(\cdot):B\mapsto\mathbb{R}$ such that,
\begin{equation*}
    E\{N_{\mathbf{X}}(B)\}=\int_B\rho(u)du
\end{equation*}
then $\rho(\cdot)$ is called the intensity function of $\mathbf{X}$. 
The intensity function is of key interest in point pattern analysis as $\rho(u)|du|$ is interpreted as the approximate probability that an event occurs in the infinitesimal set $du$.

Poisson point processes are one of the most fundamental and tractable spatial point process models. 
In practice, $\rho(\cdot)$ is often varying over $D$, i.e. $\mathbf{X}$ is inhomogeneous and can also depend on some spatial covariates $\bfz(u)$. In our study, we model the intensity function with a general log-linear form,
\begin{equation}\label{eq1}
    \rho(u;\bbeta)=\exp\{\bfz^{T}(u)\bbeta\}, u\in D
\end{equation}
where $\bfz(u)=\big(z_1(u),\cdots,z_p(u)\big)^{T}$ is a $p$-dimensional vector of spatial covariates associated with the spatial location $u$, and $\bbeta \in\mathbb{R}^p$ is the vector of regression parameter. 

There are several other popular parametric point process models whose marginal intensity functions take the same log-linear form as in \eqref{eq1}. The class of Cox process models is one such example. 
Let $\bLambda=\{\Lambda(u): u \in D\}$ denote a real, nonnegative valued random field. If the conditional distribution of $\mathbf{X}$ given $\bLambda$ is a Poisson process on $D$ with intensity function $\bLambda$, then $X$ is said to be a Cox process driven by $\bLambda$. Popular examples of Cox processes models include the Neyman-Scott process and the log Gaussian Cox process. See a review in Chapter 17 of \cite{gelfand2010handbook}. 
\subsection{Composite Likelihoods}

To estimate  $\bbeta$ in \eqref{eq1}, one commonly used method is to construct unbiased estimating equations and obtain estimators by maximizing the corresponding composite likelihoods. 
 The Poisson based composite log-likelihood function \citep{waagepetersen2007estimating} and the logistic based composite log-likelihood function \citep{baddeley2014logistic} have been used widely in the literature, which are respectively given by
\begin{equation}\label{eq:PL0}
\ell_{\text{PL}}(\bbeta)=\sum_{i=1}^m \log\rho(u_i;\mathbf{\bbeta})-\int_D\rho(u;\mathbf{\bbeta})du
\end{equation}
\begin{equation}\label{eq:LRL0}
\ell_{\text{LRL}}(\bbeta)=\sum_{i=1}^m \log(\frac{\rho(u_i;\mathbf{\bbeta})}{\delta(u_i)+\rho(u_i;\mathbf{\bbeta})})-\int_D\delta(u)\log(\frac{\rho(u;\mathbf{\bbeta})+\delta(u)}{\delta(u)})du,
\end{equation}
where $\{u_1,\cdots,u_m\}$ denotes a set of realizations of a point process, and $\delta(u)$ is  a non-negative real-valued function. 
When point process is a Poisson process, the Poisson based composite log-likelihood function in \eqref{eq:PL0} is identical to the full log-likelihood function. 
For other point processes models, the use of 
composite likelihood can be justified by the theory of estimating functions \citep{guan2006composite}. 
It can be shown \citep[see, e.g.,][]{guan2006composite,choiruddin2018convex} that the estimators obtained by maximizing both Poisson based and logistic based composite log-likelihood are the solution to the two corresponding unbiased estimating equations for $\bbeta$. 

Nevertheless, the composite likelihood based inference produces a less efficient estimator compared with the full likelihood based estimator, due to the loss of information incurred when only using the first-order moment property of the point process. To improve its efficiency, several methods have been developed to carefully select the weights when combining composite likelihood terms \citep{guan2010weighted}.  For simplicity, we only consider the unweighted composite likelihoods in the paper, but remark that the methods can be potentially generalized to the use of weighted composite likelihoods. 
 
In practice, numerical approximations are needed for the composite likelihood inference because both the evaluations of  \eqref{eq:PL0} and \eqref{eq:LRL0} involve integral terms. For equation \eqref{eq:PL0}, \cite{berman1992approximating} developed a numerical quadrature method that employs Riemann sum approximation to the integral part. 
To implement this, the domain $D$ is partitioned into $M-m$ quadrats. 
More details on how we divide a 2-D bounded domain and a line network can be found in Appendix \ref{Appendix2}. 
The $M-m$ dummy points, denoted by $\{u_i,i=m+1,\cdots,M\}$, are then placed at the centroid of each quadrat.  The Poisson based composite log likelihood is then approximated by
\begin{equation}\label{app1}
\ell_{PL}(\bbeta)\approx\sum_{i=1}^{M}v_i\{y_i\log\rho(u_i;\bbeta)-\rho(u_i;\bbeta)\},
\end{equation}
where $\{u_i \in D$, $i=1,\cdots,M\}$ consists of the $m$ observed points and $M-m$ dummy points. $v_i$ is the quadrature weight corresponding to each $u_i$. We set $v_i=a_i/n_i$, where $n_i$ denotes the total number of observed events and dummy points in the quadrat that $u_i$ resides, and $a_i$ denotes the Lebesgue measure of the quadrat of $u_i$ such that $\sum_{i=m+1}^Ma_i=|D|$. The working response data becomes $y_i=v_i^{-1}\Delta_i$, where $\Delta_i$ is an indicator of whether point $i$ is an observation ($\Delta_i=1$) or a dummy point ($\Delta_i=0$).

The Berman-Turner approximation in \eqref{app1} often requires a great amount of dummy points, consequently incurring extra computational cost. \cite{baddeley2014logistic} showed that the estimates based on the logistic likelihood in \eqref{eq:LRL0} perform competitively with the Berman-Turner approximation using a smaller number of dummy points. The method approximates \eqref{eq:LRL0} by
\begin{equation}\label{app2}
    \ell_{LRL}(\bbeta)\approx\sum_{i=1}^m\log\frac{\rho(u_i;\bbeta)}{\delta(u_i)+\rho(u_i;\bbeta)}+
    \sum_{i=m+1}^M\log\frac{\delta(u_i)}{\delta(u_i)+\rho(u_i;\bbeta)}
\end{equation}
where the integration term is calculated by Monte Carlo integration, and the dummy points are drawn from a Poisson point process over $D$, which has an intensity function $\delta(u)$ and is independent from $\mathbf{X}$. 
Applying the Campbell's formula~\citep{moller2003statistical},
it is straightforward to show that the expectation of the second term in \eqref{app2} equals to the integral part in \eqref{eq:LRL0}. 
We follow the suggestion of \cite{baddeley2014logistic} and choose $\delta(u)=(M-m)/|D|$ in our numerical studies. 



\section{Methodology}\label{sec:model}
\subsection{Spatially Varying Coefficient Models}\label{sec:SVCI}
A traditional way to model the log-linear term of the intensity function is to treat regression coefficients as constants in space as in \eqref{eq1}. In the proposed model, we are interested in estimating a piece-wise constant intensity function in an intercept-only log-linear model or detecting clustering patterns in $\bbeta$ when covariates are available.  Below, we introduce a varying coefficient log-linear intensity model (SVCI) for spatial point processes via a graph regularization method. 
 
 To elaborate, suppose a set of spatial points is observed at locations $u_1,\ldots,u_m\in D$. We assume these spatial points are a realization from a point process $\mathbf{X}$ with an intensity function $\rho(u)$,  which depends on the $p$-dimensional spatial explanatory variables $\bfz(u)=\{z_1(u),\ldots,z_p(u)\}$. As an extension of the constant coefficients regression model, we assume that
the regression coefficients are spatially varying across $D$, denoted as $\bbeta(u)=\{\beta_1(u),\ldots,\beta_p(u)\}^T$. 
The spatially varying coefficient models inherit the simplicity and easy interpretation of the traditional log-linear model in (\ref{eq1}), yet they still enjoy great flexibility that allows practitioners to investigate locally varying relationships among variables. 

Let $\bbeta_{k}= \big(\beta_k(u_1),\ldots,\beta_k(u_m)\big)^{T}$ denote the vector of regression coefficients associated with the $k$-th covariate, for $k=1,\ldots, p$. 
We assume that each $\bbeta_k$ has its own spatially clustered pattern and is piece-wise constant on $D$, that is, the coefficients are homogeneous in the same spatial cluster and varying across different clusters.  In many spatial applications involving point patterns including traffic accidents, crime locations,  pick-up/drop-off locations of Taxi trips, to name a few, it is desired to consider spatially contiguous clustering configurations such that only adjacent locations are clustered together. This way, the practitioners can detect discontinuities across boundaries and easily interpret the clusters as local regions to facilitate subsequent regional analysis. 

Before introducing our regularization method, we formally define spatially contiguous cluster of points using the notion of connected components in graph theory. Consider an undirected graph denoted as $\mathbb{G}=(\mathbb{V},\mathbb{E})$, where $\mathbb{V}=\{u_i\in D,i=1,\ldots,n\}$ is the set of vertices, and $\mathbb{E}$ is the edge set consisting of a subset of $\{(u_i,u_j):u_i,u_j\in\mathbb{V}\}$.  
In graph theory, a graph $\bbG$ is said to be connected if for any two vertices there exists a path between them. A subgraph $\bbG_{s}$ is called a connected component of $\bbG$ if it is connected and there is no path between any vertex in $\bbG_s$ and any vertex in $\bbG \setminus \bbG_s$, i.e., the difference between sets $\bbV$ and $\bbV_s$. Now we can define spatially contiguous clusters as the connected components of a graph $\bbG$. As a result, a spatially contiguous partition of $\bbV$ is defined as a collection of disjoint connect components such that the union of vertices is $\bbV$. 

This motivates us to construct a graph based regularization model, which permits contiguous cluster identifications of regression coefficients for each covariate in the log-linear point process model. Let $\bbeta_{k}^{*}=\big(\beta_k(u_{m+1}),\ldots,\beta_k(u_M)\big)^{T}$, for $k=1,\ldots, p$, denote the vector of regression coefficients at the dummy points associated with the $k$-th covariate.  
Denote the vector of the stacked regression coefficients at both the observed and dummy points by $\mathbf{\bbeta}=(\bbeta_{1}^{T},\bbeta_{1}^{*,T},\ldots,\bbeta_{p}^{T},\bbeta_{p}^{*,T})^{T}$.
We estimate $\bbeta$ by minimizing the penalized negative composite log likelihood objective function:
 
\begin{equation}\label{eq4}
Q(\bbeta)=-\frac{1}{|D|}\tilde{\ell}_c(\bbeta)+\sum_{k=1}^p\sum_{(i,j)\in\mathbb{E}}P_{\lambda}(\beta_k(u_i)-\beta_k(u_j))
\end{equation}
where $\tilde{\ell}_c(\beta)$ is either the approximation of the Poisson based composite log-likelihood or the logistic regression-based composite log-likelihood function with the following expressions: 

\begin{equation}\label{eq:PL}
\tilde{\ell}_{PL}(\bbeta) = \sum_{i=1}^{M}v_i\{y_i\log\rho(u_i;\bbeta(u_i))-\rho(u_i;\bbeta(u_i))\},
\end{equation} 
\begin{equation}\label{eq:LRL}
    \tilde{\ell}_{LRL}(\bbeta)= \sum_{i=1}^m\log\frac{\rho\big(u_i;\bbeta(u_i)\big)}{\delta(u_i)+\rho\big(u_i;\bbeta(u_i)\big)}+
    \sum_{i=m+1}^M\log\frac{\delta(u_i)}{\delta(u_i)+\rho\big(u_i;\bbeta(u_i)\big)}
\end{equation}

The second term in the objective function \eqref{eq4} adds a graph pairwise fused regularization to the negative composite log-likelihood function.   
$\bbE$ is the edge set of a graph $\bbG$, and $(i,j)\in\mathbb{E}$ implies that there is an edge in $\mathbb{E}$ connecting the points at $u_i$ and $u_j$.  
$P_{\lambda}(\cdot)$, a non-negative function tuned by parameter $\lambda$, penalizes the pairwise difference of regression coefficients whose corresponding locations are connected by an edge in $\mathbb{E}$. One popular choice is the $L_1$-penalty,
\begin{equation*}
P_{\lambda}(t)=\lambda\Vert t\Vert_1
\end{equation*}
which is often referred to as the graph fused lasso penalty in the literature \citep{tibshirani2011solution,arnold2016efficient,li2019spatial}. The $L_1$ penalty encourages sparsity in  the pairwise differences between the coefficients of edge-connected locations. As a result, the edges in the graph can be classified into 
a set that corresponds to the non-zero elements of $|\beta_k(u_i)-\beta_k(u_j)|$, and another set that corresponds to the zero elements of $|\beta_k(u_i)-\beta_k(u_j)|$. 
This naturally leads to a piece-wise constant estimate of $\bbeta_k$ for each covariate and hence a well defined spatially contiguous partition of the vertices for each covariate.  $\lambda$ is a non-negative tuning parameter that determines the strength of penalization and ultimately influences the estimates of clustered. To make a proper choice of the values of them, we use the Bayes information criterion (BIC) to select an optimal value of $\lambda$ \citep{choiruddin2021information}. Specifically, $BIC = -2\tilde{\ell}_{c} + df \log m$, where $\tilde{\ell}_c$ is the approximated composite log likelihood as in \eqref{eq:PL} and \eqref{eq:LRL}, $m$ is the number of observations, and $df$ is the degree of freedom of $\hat{\bbeta}$. Following \cite{tibshirani2011solution}, $df$ is estimated by the summation of the number of clusters for each regression coefficient $\bbeta_k$.

We remark that there are other choices of sparsity inducing penalty functions, including adaptive lasso \citep{zou2006adaptive}, smoothly clipped absolute deviation \citep[SCAD,][]{fan2001variable}, and minimax concave penalty \citep[MCP,][]{zhang2010nearly}.
And there are other criteria for tuning parameter selection, including Akaike information criterion (AIC), generalized cross-validation  \citep[GCV,][]{golub1979generalized}, and extended Bayesian information criterion  \citep[EBIC,][]{chen2012extended}.
In this paper, we choose to use Lasso together with BIC to demonstrate the utility of our method for its computational simplicity. The method can adopt other forms of penalty functions and model-selection criteria which may further improve its performance.

The selection of edge set $\mathbb{E}$ is a key ingredient in our SVCI model by playing two important roles. First, the corresponding graph $\mathbb{G}$ reflects the prior assumptions about the spatial structure and the contiguous constraint of the regression coefficients. In particular, we rely on $\mathbb{G}$ to incorporate the relational information among observations on complex constrained domains so that we can relax the Euclidean assumption. Second,  as we will explain in Section \ref{sec:Computation},  the computation speed and storage complexity of the optimization algorithm are largely determined by the structure of $\bbG$. We seek to construct a graph fused lasso regularization to achieve a good balance between model accuracy and computational efficiency.  

For point patterns on a bounded observation domain, one natural choice is to construct a nearest neighbor graph that connects each vertex with its $k$ nearest neighbors ($K$-NN) or neighbors within a certain radius ($r$-NN). In practice, the number of neighbors in $K$-NN or the radius in $r$-NN needs to be chosen with care to guarantee that $\bbG$ is a connected graph. 
It is known in machine learning literature \citep[see, e.g.,][]{shaw2009structure} that $K$-NN graphs can effectively preserve the intrinsic manifold structure of the data. Another approach is the Delaunay triangulation \citep{lee1980two}, which constructs triangles with a vertex set  such that no vertex is inside the circumcircle of any triangle. In practice, edges longer than a certain threshold are removed to ensure the spatial proximity of neighboring vertices.  Triangular graphs have also shown their capabilities in preserving complex topological structures of the data. See \cite{lindgren2011explicit,mu2018estimation} for examples.  
Moreover, when a graph has certain simple structures such as a chain or a tree graph, several recent work~\citep{padilla2018dfs,li2019spatial} showed that these simple graph structures can be utilized to design efficient algorithms to solve the graph fused lasso problem.  This motivates us to adopt a similar strategy to replace the original graph by a minimum spanning tree graph, defined as a subgraph that connects all vertices with no cycles and with minimum total edge weights. We will investigate and compare the performance of the proposed SVCI model with different types of graphs in the numerical studies in Section \ref{sec:simulation} and Appendix Section A3. 
 
\begin{figure}
\begin{center}
\includegraphics[width=3in]{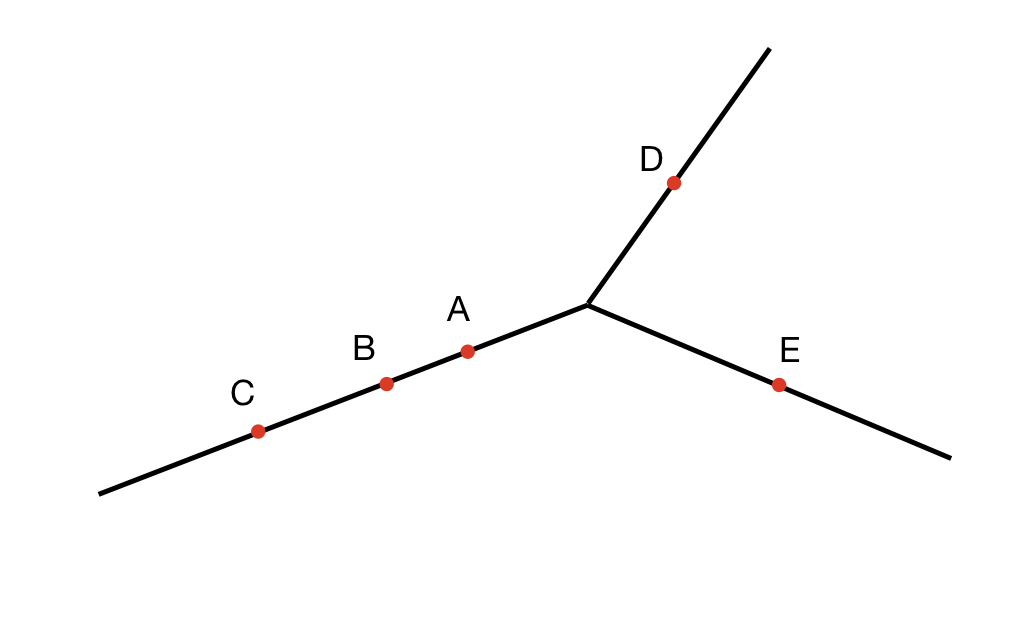}
\end{center}
\caption{A simple illustration of connections in linear networks}
\label{illu1}
\end{figure}

For point patterns on a linear network, we use an edge set that only connects pairs which are natural neighbors. To illustrate how we define natural neighbors,  we provide a simple example of a linear network (black segments) and 5 spatial points (red nodes) near an intersection in Figure~\ref{illu1}. For any interior point such as point $\mathtt{B}$, defined as a point where there exists one other point on each side of the same line, we connect it with its two adjacent points $\{\mathtt{A},\mathtt{C}\}$. For any boundary point such as point $\mathtt{A}$, defined as a point where there is no other point on the path between it and the intersection point, we connect it with $\{\mathtt{B},\mathtt{D},\mathtt{E}\}$, i.e., its adjacent interior point on the same line and its adjacent boundary points on the other lines that cross the same intersection.

\subsection{Computation}\label{sec:Computation} 
Once we construct the edge set $\mathbb{E}$, we rewrite the objective function in \eqref{eq4} in matrix form and obtain the estimate of $\bbeta$ by solving the following fused lasso optimization problem:
\begin{equation}\label{eq8}
\hat{\bbeta}=\argmin_{\bbeta\in\mathbb{R}^{^{pM} }}\{-\frac{1}{|D|}\tilde{\ell}_{c}(\bbeta)+\lambda\sum_{k=1}^p\Vert\bfH\bbeta_k\Vert_1\},
\end{equation}
where $\mathbf{H}$ is an $m^{\prime\prime}\times M$ incidence matrix corresponding to the edge set $\mathbb{E}$ with $m^{\prime\prime}$ edges. Specifically, for the $l$-th edge of $\mathbb{E}$ connecting vertices $u_i$ and $u_j$, the penalty term $|\beta_k(u_i)-\beta_k(u_j)|$ is represented as $|\mathbf{H}_l\bbeta_k|$, where $\mathbf{H}_l$ is the $l$-th row of $\mathbf{H}$ and contains only two nonzero elements; 1 at the $i$-th index and $-1$ at the $j$-th.

The path following type of algorithms \citep{arnold2016efficient}
and alternating direction methods of multipliers \citep[ADMM,][]{boyd2011distributed} have been
developed to solve graph fused lasso problems. However, the computation of these algorithms can be expensive for a general graph with a large number of nodes. Recall the number of nodes in our graph is the summation of the numbers of observations and dummy points, typically a large number in practice. It is, therefore, challenging to directly apply these conventional algorithms for the implementation of our model.

It is noted that the two approximated log composite likelihood functions in \eqref{eq:PL} and \eqref{eq:LRL} coincide with the forms of the log likelihood function of a weighted Poisson linear regression and a logistic linear regression, respectively, both of which are concave functions of $\bbeta$. Below, we propose to combine the proximal gradient method and the alternating direction method of multipliers to solve the convex optimization problem in \eqref{eq8}. In particular, we take advantage of the structures of our selected spatial graphs
to speed up computation.

Specifically, with the current estimate of the parameters being $\bbeta^{(t)}$, we follow the proximal gradient method \citep{beck2009fast} to update the value of $\bbeta$ iteratively by solving:
\begin{equation}\label{Eq:max_f}
       \bbeta^{t+1} = \argmin_{\bbeta}   \frac{1}{2}\left\Vert \bbeta - \mathbf{R}^{(t)}\right\Vert_{2}^2 + \dfrac{\lambda}{L}\sum_{k=1}^p\Vert\bfH\bbeta_k\Vert_1,
\end{equation}
where $L$ is the local Lipschitz constant of $-\frac{1}{|D|}\tilde{\ell}_c(\bbeta^{(t)})$, $\nabla{\tilde{\ell}_{c}}(\bbeta^{(t)})$ is the first derivative of $\tilde{\ell}_{PL}(\bbeta)$ or $\tilde{\ell}_{LRL}(\bbeta)$ evaluated at $\bbeta^{(t)}$ for the Poisson based or logistics regression log likelihood respectively, and $\mathbf{R}^{(t)} = \bbeta^{(t)} + (1/L)\frac{1}{|D|}\nabla{\tilde{\ell}_{c}}(\bbeta^{(t)})$. We can choose $L$ to be the maximum eigenvalue of the Hessian matrix of $-\frac{1}{|D|}\tilde{\ell}_c(\bbeta)$ evaluated at $\bbeta^{(t)}$.

Now the optimization at each iteration boils down to 
solving \eqref{Eq:max_f}, which we propose to use the ADMM algorithm \citep{wahlberg2012admm}.  By introducing auxiliary variables $\btheta=\{\btheta_1,\cdots,\btheta_p\}$, Equation \eqref{Eq:max_f}
is equivalent to:
\begin{equation*}
\argmin_{\bbeta,\btheta} \frac{1}{2}\left\Vert \bbeta - \mathbf{R}^{(t)}\right\Vert_{2}^2+\dfrac{\lambda}{L}\sum_{k=1}^p\Vert\btheta_k\Vert_1
\qquad s.t.\quad\btheta_k=\mathbf{H}\bbeta_k,\quad\forall  k=1,\cdots,p
\end{equation*}
Its augmented Lagrangian function is:
\begin{equation*}
\frac{1}{2}\left\Vert \bbeta - \mathbf{R}^{(t)}\right\Vert_2^2+\dfrac{\lambda}{L}\sum_{k=1}^p\Vert\btheta_k\Vert_1+\frac{\gamma}{2}\sum_{k=1}^p\Vert\mathbf{H}\bbeta_k-\btheta_k \Vert^2_2+\gamma\sum_{k=1}^p\mathbf{u}^T_k(\mathbf{H}\bbeta_k-\btheta_k)
\end{equation*}
where $\mathbf{u}=\{\mathbf{u}_1,\cdots,\mathbf{u}_p\}$ are Lagrangian multipliers, and $\gamma$ is a penalty parameter. ADMM alternately optimizes $\{\bbeta,\btheta,\mathbf{u}\}$ by solving the following three subproblems:

\begin{equation*}
\left\{
\begin{aligned}
\bbeta^{(t+1)}&=\arg\min_{\bbeta}\{\left\Vert \bbeta - \mathbf{R}^{(t)}\right\Vert_{2}^2+\gamma\sum_{k=1}^p\Vert\mathbf{H}\bbeta_k-\btheta^{(t)}_k+\mathbf{u}_k^{(t)} \Vert_2^2\},\\
\btheta_k^{(t+1)}&=\arg\min_{\btheta_k}\{\frac{\lambda}{L}\Vert\btheta_k\Vert_1+\frac{\gamma}{2}\Vert\mathbf{H}\bbeta^{(t+1)}_k-\btheta_k+\mathbf{u}^{(t)}_k \Vert_2^2\},\quad k=1,\cdots,p\\
\mathbf{u}^{(t+1)}_k&=\mathbf{u}^{(t)}_k+\mathbf{H}\bbeta^{(t+1)}_k-\btheta^{(t+1)}_k,\quad k=1,\cdots,p
\end{aligned}
\right.
\end{equation*}
where $t$ denotes the $t$-th iteration.

The above sub optimization problems have the following analytical results:
\begin{equation*}
\left\{
\begin{aligned}
&\bbeta_k^{(t+1)}:(\bfI_{n}+\gamma\mathbf{H}^T\mathbf{H})^{-1}[\mathbf{R}^{(t)}_k+\gamma\mathbf{H}^T(\btheta_k^{(t)}-\mathbf{u}_k^{(t)})]\\
&\theta_k^{(t+1)}:\mathcal{S}(\mathbf{H}\bbeta^{(t+1)}_k+\mathbf{u}_k^{(t)};\frac{\lambda}{L\gamma}),\\
&\mathbf{u}^{(t+1)}_k=\mathbf{u}^{(t)}_k+\mathbf{H}\bbeta^{t+1}_k-\btheta^{t+1}_k,\\ 
\end{aligned}
\right.
\end{equation*}
for each $k=1,\cdots,p$, where $\mathcal{S}(z,\lambda)$ is the soft-thresholding operator, and $\mathbf{R}^{(t)}_k = \bbeta^{(t)}_k + (1/L)\frac{1}{|D|}\nabla{\tilde{\ell}_{c}}(\bbeta^{(t)}_k)$. It is noted that the above optimization steps are separable for the parameters associated with each $k$, and hence can be conveniently solved in a parallel fashion. In addition, under our choice of the spatial graphs, the graph Laplacian matrix $\bfH^{T}\bfH$ is a sparse matrix. 
As a result, the update of $\bbeta^{(t)}$ only involves the linear solver of the sparse matrix $(\bfI_{n}+\gamma\mathbf{H}^T\mathbf{H})^{-1}$, whose sparse Cholesky factorization can be pre-computed efficiently using \texttt{R} package \texttt{Matrix}. We iterate between (8) and the above ADMM steps until convergence.

\subsection{Theoretical analysis}\label{sec:theory}

In this section, we adopt an increasing domain framework (see Assumption 1 below for details) and investigate the rates of convergence for our estimators when the expansion rate $n$ goes to infinity. For simplicity, we only present the asymptotic results for the regularized unweighted Poisson based composite likelihood estimator. Similar results and proofs hold for the regularized logistic based composite likelihood estimator.  

We first define notations needed to establish the theoretical results below: 
\begin{itemize}
\item 
$\hat{\bbeta}_{n}$ denotes $\hat{\bbeta}$ obtained over $D_{n}$ using \eqref{eq:PL}, and $\bbeta^{0}$ denotes the true parameter value.
\item $\rho^{(k)}\left(u_{1}, \ldots, u_{k}\right)\equiv \lim _{\substack{\left|d u_{i}\right| \rightarrow 0 \\ i=1, \ldots, k}}\left[\dfrac{E\left\{N\left(d u_{1}\right) \cdots N\left(d u_{k}\right)\right\}}{\left|d u_{1}\right| \cdots\left|d u_{k}\right|}\right]$ denotes the $k$-th order intensity function. 
\item $
G_{k}\left(u_{1}, \ldots, u_{k}\right)=\lim _{\substack{\left|d u_{i}\right| \rightarrow 0 \\ i=1, \ldots, k}}\left[\frac{\operatorname{cum}\left\{N\left(d u_{1}\right), \ldots, N\left(d u_{k}\right)\right\}}{\left|d u_{1}\right| \cdots\left|d u_{k}\right|}\right]
$
denotes the cumulant function of a point process describing the dependence among points $u_1,\ldots,u_k$, where we denote the notation $\operatorname{cum}\left(N_{1}, \ldots, N_{k}\right)$ as the coefficient of $i^{k} t_1 t_2 \cdots t_{k-1}t_k$ in the Taylor series expansion of $\log \left[E\left\{\exp \left(i \sum_{j=1}^{k} N_{j} t_{j}\right)\right\}\right]$ at the origin.  
\item The strong mixing coefficient is defined as
\begin{equation*}
\begin{split}
\alpha(p ; k)\equiv \sup \{|P(A_{1} \cap A_{2})- & P(A_{1}) P(A_{2})|:  A_{1} \in \mathscr{F}(\tilde{E}_{1}), A_{2} \in \mathscr{F}(\tilde{E}_{2}), \\ 
& \tilde{E}_{2}=\tilde{E}_{1}+s, |\tilde{E}_{1}|=|\tilde{E}_{2}| \leq p, d(\tilde{E}_{1}, \tilde{E}_{2}) \geq k \},
\end{split}
\end{equation*} where $\mathscr{F}$ is the $\sigma$-algebra generated by $\bfX \cap \tilde{E}_{i}, i=1,2$, $d\left(\tilde{E}_{1}, \tilde{E}_{2}\right)$ is the minimal distance between sets $\tilde{E}_{1}$ and $\tilde{E}_{2}$,  and the supremum is taken over all compact and convex subsets $\tilde{E}_{1} \subset \mathbb{R}^{2}$ and over all $u \in \mathbb{R}^{2}$.
\item $U_{n}\left(\bbeta\right))=-\ell(\bbeta)/|D_n|$ denotes the scaled negative Poisson composite log-likelihood function. $U_{n}^{(1)}\left(\bbeta\right)$ and $U_{n}^{(2)}\left(\bbeta\right)$ denote the first and second derivatives of $U_{n}\left(\bbeta\right)$ respectively. 

\item \begin{eqnarray*} 
\mathbf{B}_{n}\left( \boldsymbol{\beta}\right) & = & \int_{D_{n}}  \mathbf{z}(u) \mathbf{z}(u)^{\top} \rho\left(u ; \boldsymbol{\beta}\right) \mathrm{d} u \\
\mathbf{C}_{n}\left( \boldsymbol{\beta}\right) &= & \int_{D_{n}} \int_{D_{n}}  \mathbf{z}(u) \mathbf{z}(v)^{\top}\{g(u, v)-1\} \rho\left(u ; \boldsymbol{\beta}\right) \rho\left(v ; \boldsymbol{\beta}\right) \mathrm{d} v \mathrm{d} u
\end{eqnarray*}
Below, we use $\bfB_n$ and $\bfC_n$ to denote for $\bfB_n(\bbeta^0)$ and $\bfC_n(\bbeta^0)$ respectively.
\item Let $H$ be the direct sum of $p$ incidence matrices $\bfH$.
Let $H^{\dagger}$ be the the Moore-Penrose pseudo inverse of $H$,  $P_{H}=H^{\dagger}H$ be the projection matrix onto the row space spanned by $H$, and  $P_{H^{\perp}}=I-P_{H}$. 
 
\end{itemize}
 
Our asymptotic results rely on the following regularity conditions as $n \rightarrow \infty$:
\begin{assumption}\label{as:1}
For every $n \geq 1, D_{n}=\{n e: e \in D_{1}\}$ where $D_{1} \subset [0,1]\times [0,1]$ and $|D_1| = \text{Const.}$ 
Assumption 1 states that we consider an increasing domain asymptotic where both coordinates for each interior points of $D_n$ expand by $n$. 
\end{assumption}
\begin{assumption}\label{as:2}   
$$
\sup _{p} \frac{\alpha(p ; k)}{p}=\mathrm{O}\left(k^{-\epsilon}\right), \quad \text { for some } \epsilon>2
$$
Assumption 2 is the strong mixing coefficient condition such that for any two fixed set, dependence between them decays to 0 at a polynomial rate of the intersect distance $k$.  
\end{assumption}
\begin{assumption}\label{as:3}
 The first-order intensity function $\rho(u ; \bbeta)$ is bounded below from 0,  $\rho^{(2)}(u,u' ; \bbeta)$ is bounded and continuous with respect to $\bbeta$, and $\sup _{u_{1}} \int_{D_{n}} \cdots \int_{D_{n}}\left|G_{k}\left(u_{1}, \ldots, u_{k}\right)\right| d u_{2} \cdots d u_{k}<C$ for $k=2,3,4$.
\end{assumption}
\begin{assumption}\label{as:4}
$\liminf _{n \rightarrow \infty} \nu_{\min }\left(\left|D_{n}\right|^{-1}\left\{\mathbf{B}_{n}+\mathbf{C}_{n}\right\}\right)>0$, where  $\nu_{\min}(\cdot)$ denotes the smallest eigenvalue of a matrix.
\end{assumption}

\begin{assumption}\label{as:5}
 There exists a $\delta$-neighborhood of $\bbeta^0$ denoted as $\mathcal{N}_{\delta}(\bbeta^0)$, such that $\widehat{\bbeta}_n\in \mathcal{N}_{\delta}(\bbeta^0)$ and 
$\liminf _{n \rightarrow \infty} \nu_{\min }\big( \left|D_{n}\right|^{-1} \mathbf{B}_{n}(\boldsymbol{\beta})\big)>0$, for any $\bbeta \in \mathcal{N}_{\delta}(\bbeta^0)$.
\end{assumption}

Assumptions 1-4 are commonly adopted in the asymptotic theories for point process models. We adopt these assumptions to establish the asymptotic normality result for $U_n^{(1)}(\bbeta^0)$. Assumption 5 is adopted for second order Taylor expansion of the composite likelihood function. These assumptions are needed to derive the error bound for $\widehat{\bbeta}_n$. We then follow similar ideas as the basic inequality in Lasso \citep{buhlmann2011statistics} and the projection argument in \cite{wang2016trend} to derive the following error bound.

\begin{theorem}\label{eq:theorem}
Let $M^{*}$ denote the maximum $L_2$ norm of the columns of $(\bfB_1+\bfC_1)^{1/2}P_{{H}^{\perp}}H^{\dagger}$,   
under Assumptions 1 to 5, for a tuning parameter $\lambda= O(M^{*} |D_n|^{-1/2} \sqrt{\log Mp})$, 
with probability tending to 1 as $n \to \infty$, we have
\begin{equation*}
    \left\|\widehat{\bbeta}_n-\bbeta^{0}\right\|_{2}^{2} =O_{\mathbb{P}}(M^{*} |D_n|^{-1/2} 
    \sqrt{\log Mp}\left\|H \bbeta^{0}\right\|_{1})
\end{equation*} 

\end{theorem}
The proof of Theorem \ref{eq:theorem} is provided in the Appendix \ref{Appendix1}. 

Finally, we demonstrate in Corollary 1 how the estimation error rate can be used to guide the detection of clusters in practice. Define $\bfH_j$ as the $j$-th row of $\bfH$, and $\mathcal{I}_k=\left\{j: \bfH_j \bbeta^0_k \neq 0\right\}$ as the set consisting of the edges that connect the points with different coefficient values for the $k$-th covariate. We have the following corollary based on Theorem 1:

\begin{corollary}\label{eq:corollary}
If $M^{*}\|H \bbeta^{0}\|_{1} = o(|D_n|^{1/2}/\sqrt{\log Mp})$,  
under the Assumptions in Theorem~\ref{eq:theorem}, there exists $\delta >0$ such that
 $|\bfH_j\hat{\bbeta}_k| < \delta \Leftrightarrow j \notin \mathcal{I}_k$
with probability tending to 1 as $n \rightarrow \infty$.
\end{corollary}

\section{Simulation Studies}\label{sec:simulation} 
In this section, we conduct simulation studies to investigate the performance of the SVCI model. We design two different data generation scenarios:
\begin{itemize} 
    \item Scenario 1: Point patterns are generated from a Poisson point process on a planar window, where the log intensity is a linear function of an intercept and two covariates with clustered regression coefficients, i.e.,
    $\rho(u;\bbeta(u))=\exp\{\beta_0(u)+z_1(u)\beta_1(u)+z_2(u)\beta_2(u)\}$.
    \item Scenario 2: Point patterns are generated from a Poisson point process on a linear network. We consider two sub-scenarios: (a) The log intensity function is piece-wise constant, i.e., $\rho(u;\bbeta(u))=\exp\{\beta_0(u)\}$; (b) The log intensity is a linear function of an intercept and two covariates with clustered regression coefficients as in Scenario 1.
\end{itemize}
 
In Scenario 1, we focus on examining the performance of our method under different model choices, including the choice of graphs used in the graph fused lasso penalty and the choice between the Poisson likelihood based SVCI (SVCI-PL) and the logistic  likelihood based SVCI (SVCI-LRL). 
For comparison studies, to the best of our knowledge, there are very limited existing methods available for spatially clustered coefficient log linear point process models on complex domains as reviewed in the Introduction, except for the simple case of an intercept-only log-linear model.
As such, Scenario 2(a) is included so that we can compare SVCI with the nonparametric kernel density estimation method on a linear network (KDE.lpp) proposed in \cite{mcswiggan2017kernel}, the fast KDE method (KDEQuick.lpp) in \cite{rakshit2019fast}, and the resample-smoothed Voronoi intensity estimation method (Voronoi.lpp) in \cite{moradi2019resample}.
For the case that has spatial covariates as in Scenario 2.(b), the comparison is made with the LGCP model \citep{moller1998log}, in which the inhomogeneity of the intensity function is modeled by a latent spatial Gaussian process random effects model.

 

Given the estimator $\hat{\bbeta}$ defined in \eqref{eq8},  we can predict the coefficients at any given new location $u\in D\backslash\{u_1,\cdots,u_M\}$ according to $\hat{\bbeta}(u)=\sum_{i=1}^M\mathbf{1}_{\{u_i\in\mathcal{N}_K(u)\}}\hat{\bbeta}(u_i)$/K, where $\mathcal{N}_K(u)$ denote the $K$ nearest neighbors of $u$. To quantify the performance of parameter estimation, we evaluate the estimation accuracy of $\beta_k(u)$ by the mean integrated squared error ($\text{MISE}_{\beta}$, \cite{davis1977mean}), defined as
\begin{equation*}
\text{MISE}_{\beta}=\frac{1}{p|D|}\sum_{k=1}^p\int_D(\beta_k(u)-\hat{\beta}_k(u))^2du
\end{equation*}

We implement our methods in \texttt{R} (see \href{https://github.com/LihaoYin/SVCI}{https://github.com/LihaoYin/SVCI}). The data generations are done using the \texttt{R} package \texttt{spatstat}  \citep{baddeley2014package}. 
The competing KDE.lpp and KDEQuick.lpp methods are implemented using    \texttt{density.lpp} and \texttt{densityQuick.lpp} in \texttt{R} package \texttt{spatstat}, respectively. Voronoi.lpp is implemented using 
\texttt{densityVoronoi.lpp} in \texttt{R} package \texttt{stlnpp}. The competing LGCP method is implemented in \texttt{R} using the \texttt{lgcp} function provided in \texttt{geostatsp} \citep{geostatsp}.  All computations were performed on a Mac Pro with 2.4 GHz Intel Core i7 laptop with 8GB of memory.



\subsection{Simulation Scenario 1}\label{sc1}
\begin{figure}
\begin{center}
\includegraphics[width=6.2in]{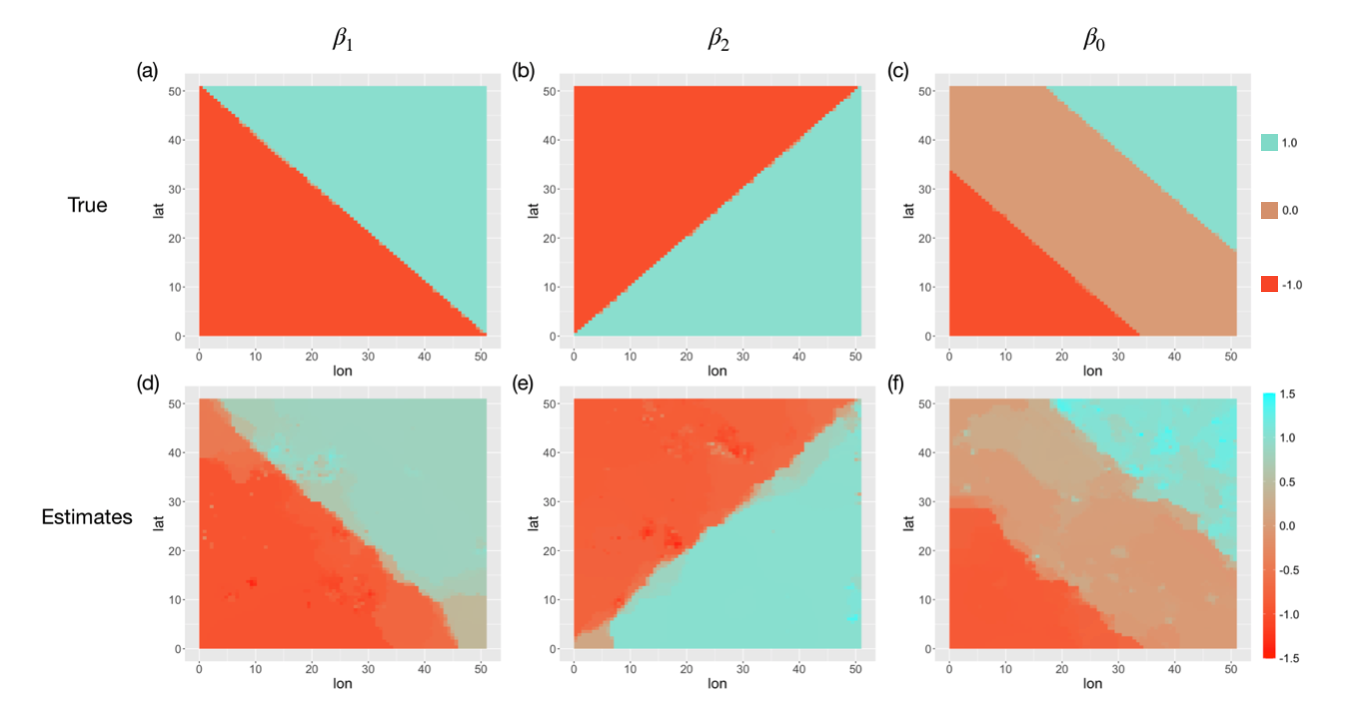}
\end{center}
\caption{The top panel (a-c): the true coefficients $\bbeta_0$, $\bbeta_1$ and $\bbeta_2$ in Scenario 1; The bottom panel (d-f): the estimated coefficients from one simulation using the $5$NN graph when $m=2400$, $\mathtt{nd}^2=n$}
\label{eg0}
\end{figure}

In Simulation Scenario 1, we consider a spatial 2D window $D= [0,R]^2\subset\mathbb{R}^2$, where the true regression coefficients in the log-intensity function are assumed to have clustering patterns as shown in the top panel of Figure~\ref{eg0}. 
We simulate the two covariates $\{z_1(u)\}$ and $\{z_2(u)\}$ from two independent realizations of a spatial GP with mean zero and an isotropic exponential covariance function taking the form of $\text{Cov}\{z_k(u),z_k(v)\}=\sigma^2\exp(-\Vert u-v\Vert/\phi)$, $k=1,2$, $u,v\in[0,R]^2$, where the range parameter $\phi=0.3R$ corresponding to a moderate spatial correlation setting. 


Under one chosen fixed coefficient pattern, we experiment with a range of $R$ values to simulate one realization from the Poisson point process model described in Scenario 1 such that the number of simulated points ranges from $800$ to $6000$ on average, in order to examine the performance of SVCI as the sample size increases with the expanding domain. Furthermore, we report the model performance for three different numbers of dummy points $\mathbb{nd}$: (a) $\mathtt{nd}^2<m$; (b) $\mathtt{nd}^2=m$; (c) $\mathtt{nd}^2>m$, where $m$ is the number of the observed points. 
We also compare with an LGCP model with intensity function $\log\rho(u)=z(u)^T\bbeta+\phi(u)$, where $\bbeta$ are the constant-coefficients across the domain, and $\phi(u)$ is a spatial Gaussian  process  with a zero  mean  and  a  M\'atern  correlation  function.

As discussed in Section \ref{sec:model}, the selection of connection graphs for the fused lasso penalty plays critical roles on the estimation accuracy and computation speed. In this study, we compare the performance of SVCI using three types of connection graphs, including the minimum spanning tree graph (MST), the Delaunay triangulation (DTs) and the $K$-nearest neighbor graph (K-NNs, $K$ is set to be $3,4,5$). Also see a comparison study between $K$-NNs and $r$-NNs in Appendix Section A3. We run $100$ repeated experiments of the SVCI model using each connection graph for both SVCI-PL and SVCI-LRL with a  fix number of dummy points $\mathtt{nd}^2=m$. 

\begin{figure}
\begin{center}
\includegraphics[width=6in]{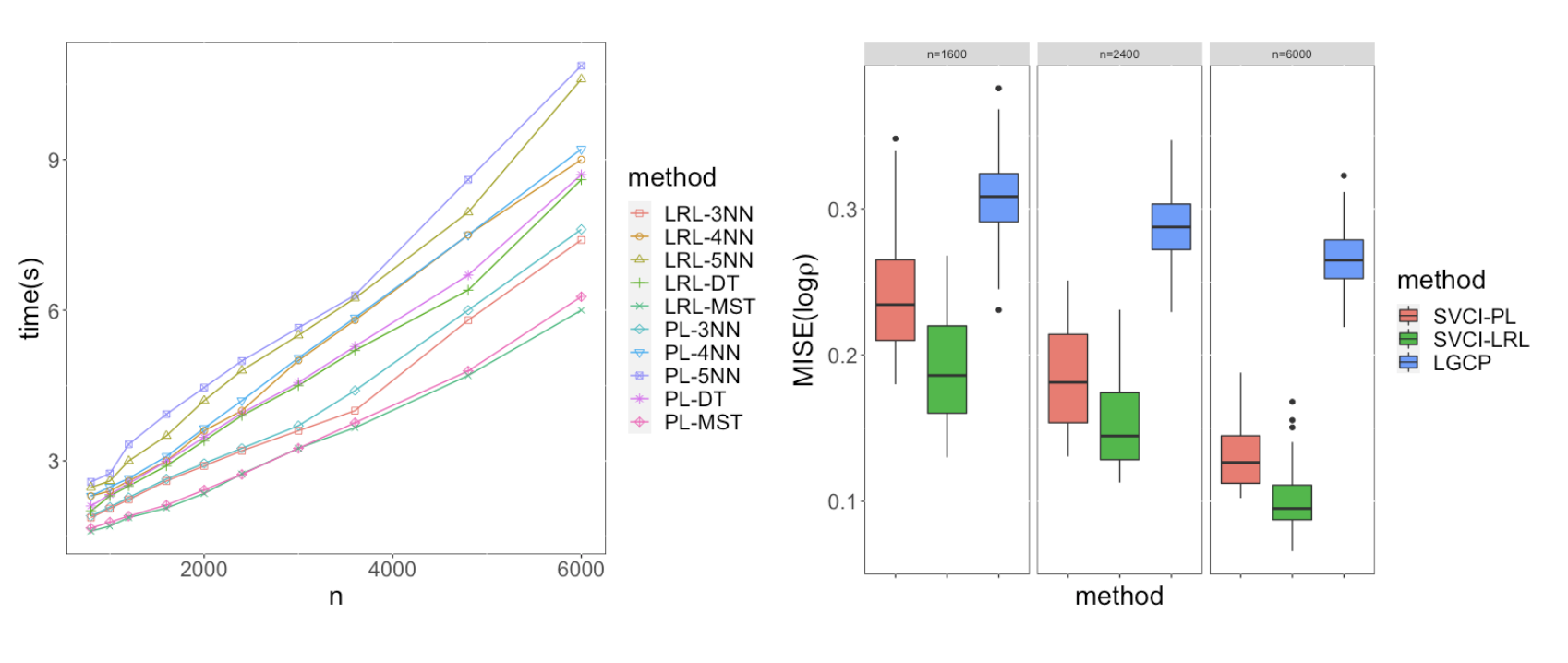}
\end{center}
\caption{ Left: Computation time to solve the optimization for one tuning parameter using different connection graphs; Right: the boxplots of $\text{MISE}_{\beta}$ for SVCI-PL, SVCI-LRL with $5$-NN graphs and LGCP. Reported results are averaged over $100$ Monte Carlo simulations under Scenario 1.}
\label{s1-2}
\end{figure}

In Table~\ref{tab1}, we report the averaged MISE of the estimated $\bbeta$ ($\text{MISE}_{\beta}$). 
There are several noticeable observations. First, a denser graph such as the $5$-NN graph produces a more accurate estimation result compared with that of a sparser graph such as the MST or $3$-NN graph. The bottom panel of Figure~\ref{eg0} illustrates an example of the estimated coefficients using the $5$-NN graph when $m=2000$, $\mathtt{nd}^2=m$, which demonstrates the capability of SVCI in capturing the cluster structure in the regression coefficients and detecting the abrupt changes across the boundaries of adjacent clusters. However, there is clearly a trade off between the estimation accuracy and computation efficiency when using different graphs; as reported in the left panel of Figure~\ref{s1-2}, the computation time (in seconds) using the $5$-NN or Delaunay triangulation graph is roughly $1.5$ times of the computation time using the MST. Second, the parameter estimation is more accurate when $m$ grows larger, as evidenced by the decreasing value of $\text{MISE}_{\beta}$.  
Finally, SVCI-LRL produces comparable results with those from SVCI-PL when $m=800$ or using the MST graph, but it notably outperforms SVCI-PL when $m\ge1600$. This is consistent with the findings in \cite{baddeley2014logistic}, which showed that for datasets with a large number of points or a highly structured point pattern, the logistic likelihood method produces a less biased estimator than its Poisson counterpart.

We further examine the performance of the SVCI model in terms of recovering the intensity function. Given $\hat{\bbeta}(u)$, we obtain the estimate of the log intensity function by $\log\hat{\rho}(u)=\bfz^T(u)\hat{\bbeta}(u)$ for $u\in D$. The right panel of Figure~\ref{s1-2} compares the $\text{MISE}$ of $\log\hat{\rho}(u)$ from SVCI-PL, SVCI-LRL and LGCP, respectively. It is noted that SVCI-LRL maintains its superior performance when predicting the intensity function in comparison with SVCI-PL. Besides, both versions of SVCI produce more accurate estimates than LGCP when estimating the intensity function with clustered regression coefficients.  

Next we examine the performance of SVCI in recovering the clusters of coefficients. Table~\ref{tab1-2} reports the Rand index for each of the SVCI estimates $\hat{\bbeta}_1$, $\hat{\bbeta}_2$ and $\hat{\bbeta}_0$ averaged over 100 simulations, using the $5$-NN graph and setting $\mathtt{nd}^2=m$. Rand index measures the proportion of pairs consisting of a true parameter and the corresponding estimated parameter that agree by virtue of belonging either to the same cluster or to different clusters.
Overall, SVCI achieves an accurate cluster recovery result, evidenced by the relatively high Rand index value ranging from 0.73 to 0.93 in all settings.  We also find that SVCI-LPL surpasses SVCI-PL in detecting spatial clusters. Finally, an interesting observation is that $\hat{\bbeta}_0$ has a lower Rand index value than both $\hat{\bbeta}_1$ and $\hat{\bbeta}_2$, which might be the consequence of having more clusters in the true function of $\bbeta_0$.

Finally we check the sensitivity of the model performance to the number of dummy points. We fix $m=1600$ and consider three different numbers of dummy points denoted by $\mathtt{nd}^2$. Table~\ref{tab2} presents the averaged $\text{MISE}_{\beta}$ and the associated computation time over $100$ simulations. For the Poisson likelihood, the default choice suggested in the \texttt{R} package \texttt{spatstat} is $\mathtt{nd}^2\approx4m$. In our experiments, however, as presented in Table~\ref{tab2}, both SVCI-PL and SVCI-LRL achieve the minimal $\text{MISE}_{\beta}$ when $\mathtt{nd}^2=60^2$, i.e. when the number of dummy points roughly equals the number of points.  Moreover, based on the results in Table~\ref{tab2}, we observe that when $\mathtt{nd}^2$ is not too large,  both SVCI-PL and SVCI-LRL seem to achieve a smaller $\text{MISE}_{\beta}$ but at a higher computation cost as $\mathtt{nd}^2$ increases. Weighing the trade-off between computation efficiency and estimation accuracy, we recommend to use $\mathtt{nd}^2\approx m$ in practice. 
 
\begin{table}[!ht]
    \caption{Scenario 1: Mean integrated squared error of $\bbeta$ ($\text{MISE}_{\beta}$) averaged over $100$ simulations for different values of $m$ with $\mathtt{nd}^2=m$, different connection graphs, and the Poisson-based SVCI-PL method and the logistic regression based SVCI-LRL method.}
    \centering
    \begin{tabular}{cc c|cc|cc|cc|cc}
    \hline\hline
    
    \multirow{2}{*}{Method}&\multicolumn{2}{c}{$m=800$}&\multicolumn{2}{c}{$m=1600$}&\multicolumn{2}{c}{$m=2400$}&\multicolumn{2}{c}{$m=3600$}&\multicolumn{2}{c}{$m=6000$} \\
    \cline{2-11}
    &PL&LRL&PL&LRL&PL&LRL&PL&LRL&PL&LRL\\
    \hline\hline
    \multicolumn{11}{c}{$\text{MISE}_{\beta}$}\\
    \hline
    MST&0.234&0.243&0.222&0.224&0.189&0.191&0.152&0.146&0.130&0.125\\
    $3$-NN&0.223&0.230&0.204&0.182&0.189&0.184&0.155&0.142&0.127&0.115\\
    $4$-NN&0.214&0.209&0.200&0.181&0.157&0.132&0.133&0.116&0.115&0.098\\
    $5$-NN&0.209&0.195&0.184&0.153&0.141&0.119&0.122&0.105&0.095&0.078\\
    DT&0.215&0.211&0.174&0.141&0.128&0.111&0.113&0.097&0.080&0.072\\
    \hline
    
    \end{tabular}
    \label{tab1}
\end{table}

\begin{table}[!ht]
    \caption{Scenario 1: Rand index of the estimates $\hat{\bbeta}_1$, $\hat{\bbeta}_2$ and $\hat{\bbeta}_0$ (averaged over $100$ Monte Carlo simulations) for SVCI-PL and SVCI-LRL for different values of $m$, using $\mathtt{nd}^2=n$ and $5$-NN connection graphs. }
    \centering
    \begin{tabular}{cc c|cc|cc|cc|cc}
    \hline\hline
    \multirow{2}{*}{}&\multicolumn{2}{c}{$m=800$}&\multicolumn{2}{c}{$m=1600$}&\multicolumn{2}{c}{$m=2400$}&\multicolumn{2}{c}{$m=3600$}&\multicolumn{2}{c}{$m=6000$} \\
    \cline{2-11}
    &PL&LRL&PL&LRL&PL&LRL&PL&LRL&PL&LRL\\
    \hline\hline
    \multicolumn{11}{c}{Rand Index}\\
    \hline
    $\hat{\bbeta}_1$&0.817&0.839&0.857&0.874&0.883&0.903&0.897&0.915&0.917&0.930\\
    $\hat{\bbeta}_2$&0.817&0.836&0.851&0.867&0.879&0.906&0.893&0.917&0.914&0.925\\
    $\hat{\bbeta}_0$&0.729&0.760&0.742&0.773&0.767&0.786&0.787&0.806&0.803&0.825\\
    \hline
    \end{tabular}
    \label{tab1-2}
\end{table}

\begin{table}[!ht]
    \centering
    \caption{Scenario 1: Comparing the $\text{MISE}_{\beta}$ (averaged over $100$ simulations) and computation time (in seconds) between SVCI-PL and SVCI-LRL for different numbers of dummy points, using $m=1600$ and $5$-NN connection graphs.}
    \begin{tabular}{ccc|cc}
         \hline\hline
         \multirow{2}{*}{dummy points}&\multicolumn{2}{c}{SVCI-PL}&\multicolumn{2}{c}{SVCI-LRL}\\
         \cline{2-5}
         &$\text{MISE}_{\beta}$&time(s)&$\text{MISE}_{\beta}$&time(s)\\
         \hline\hline
         $\mathtt{nd}^2=30^2$&0.195&1.54&0.164&1.61\\
         $\mathtt{nd}^2=40^2$&0.184&1.86&0.153&1.81\\
         $\mathtt{nd}^2=60^2$&0.182&2.38&0.151&2.30\\
         $\mathtt{nd}^2=80^2$&0.190&3.35&0.167&3.24\\
         \hline
    \end{tabular}
    \label{tab2}
\end{table}

\subsection{Simulation Scenario 2}\label{sc2}
\begin{figure}
\begin{center}
\includegraphics[width=6in]{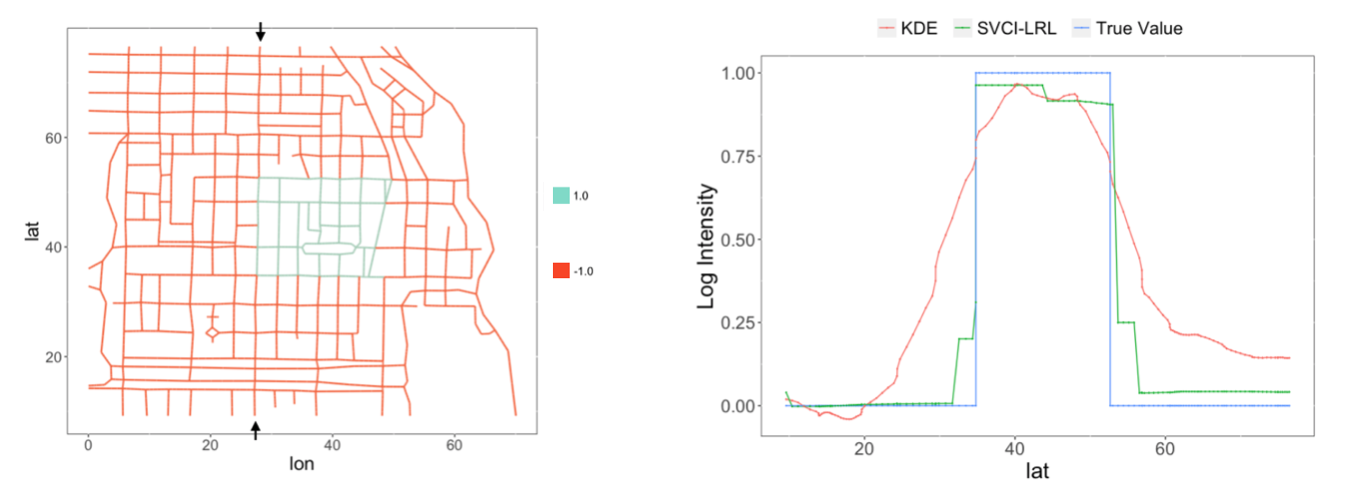}
\end{center}
\caption{Left: The true log intensity $\beta_0$ on the $\mathtt{chicago}$ network in Scenario 2(a); Right: The true and the estimated log intensity along one road segment corresponding to the line between the two black arrows in the left panel of Figure~\ref{Chi1}. }
\label{Chi1}
\end{figure}

In Scenario 2, we generate spatial points on the $\mathtt{chicago}$ linear network from the $\mathtt{R}$ package $\mathtt{spatstat}$. The network is shown in the left panel of Figure~\ref{Chi1}, which consists of the road network in an area of Chicago, USA near the University of Chicago \citep{baddeley2014package}. We bound the linear network in a window $D=R\times[0,1]^2$ and increase $R$ to expand $D$, so that the linear network that resides in $D$ grows with $D$ at the same rate to obtain an increasing number of realizations on the network.

We first consider a simplified case where there is no covariate available. We focus on the estimation of intensity function whose true value is a piece-wise constant function, that is, the intensity function $\rho(u)=\exp\{\beta_0(u)\}$,  and the true value of $\beta_0(u)$ has a clustered pattern as shown in the left panel of Figure~\ref{Chi1}. The original graph is constructed following the method described in the last paragraph of Section 3.1. The upper part of Table~\ref{tab4} presents the $\text{MISE}$ of $\log\rho$, i.e., the log intensity function for each value of $m$. In general, we obtain similar findings on the linear network as on a planar window presented in Scenario 1; $\text{MISE}$ from both the Poisson based and logistic based SVCI models show a convergence tendency as the domain expands and $m$ goes up, and the logistic likelihood based method achieves a slightly more accurate estimation than the Poisson based method with a large number of points. 
It is clear from Table~\ref{tab4} that both the SVCI-PL and SVCI-LRL models outperform the KDE based and resample-smoothed Voronoi based intensity estimation methods in almost all settings. 
Previous studies \citep{barr2010voronoi} show that KDE estimators may suffer from the problem of having substantial bias and high variance when there are abrupt changes in the intensity. 
Both SVCI and Voronoi.lpp are designed to alleviate this problem, as evidenced by their improved performance over the two KDE methods in Table 4. Nevertheless, SVCI seems to be more effective than Voronoi.lpp to capture abrupt changes or clustering patterns. 
We illustrate an example in the right panel of Figure~\ref{Chi1}, which plots the true and the estimated log intensity along a selected road segment from one simulation. It clearly shows that SVCI captures the intensity with discontinuities more efficiently than KDE.lpp. 

We also compare the computation time of each method and report the detailed results in Appendix Table A2 for various values of $m$. Taking $m=2400$ as an example, to get one estimate, KDE.lpp requires 4.95 seconds, KDEQuick.lpp requires $0.084$ seconds, and Voronoi.lpp requires $4.95$ seconds. In contrast, SVCI-LRL need $0.93$ seconds to construct the connection graph and $1.11$ seconds to get an estimate. Although SVCI is not the fastest among the compared methods, overall, its computation is still reasonable and competitive, especially considering its superior performance in intensity estimations.


\begin{figure}
\begin{center}
\includegraphics[width=6.3in]{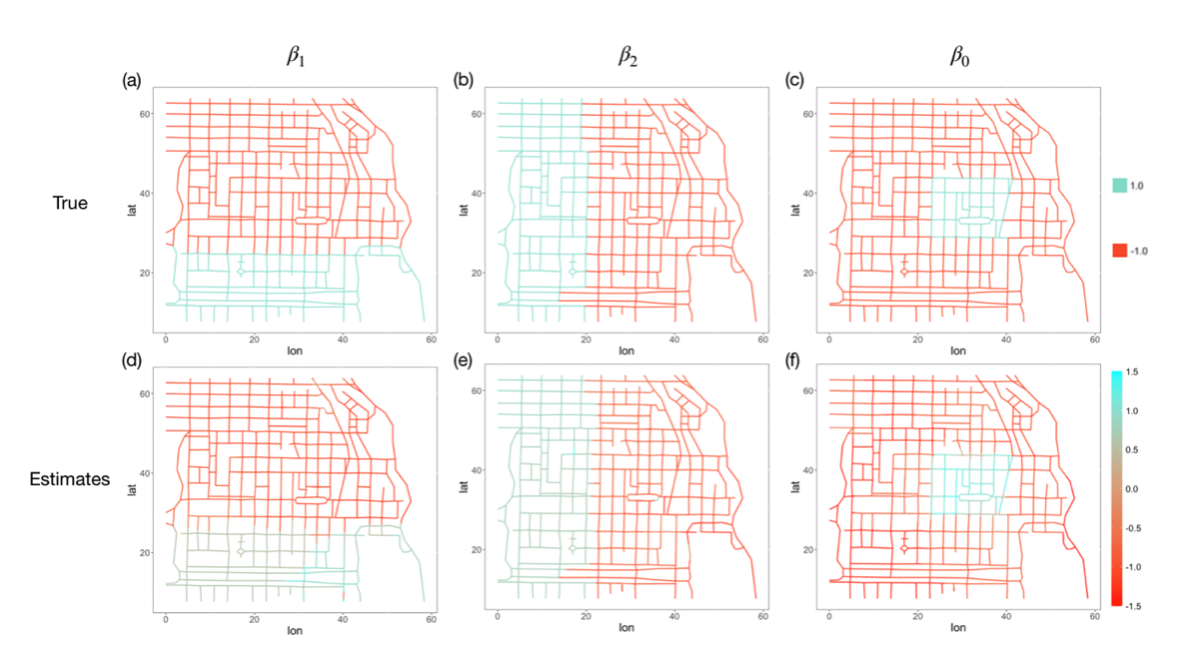}
\end{center}
\caption{The top three panels (a-c): the spatial structures of true coefficients $\bbeta_1$, $\bbeta_2$ and $\bbeta_0$ in Scenario 2(b); The upper three panels (d-f): the estimated coefficient surfaces from in one simulation using SVCI-LRL with $n=2400$.}
\label{eg2}
\end{figure}

We then consider the case with an intercept and two covariates as described in Scenario 2(b). The true regression coefficients are plotted in the subfigures (a-c) of Figure~\ref{eg2}. The subfigures (d-f) of
Figure~\ref{eg2} give the estimated coefficients from SVCI on the \texttt{chicago} network. The results demonstrate that our method is also capable of capturing the clustered coefficient patterns on a linear network. In addition, the log intensity estimation results presented in the lower part of Table~\ref{tab4} are in general consistent with the findings presented in Scenario 1 and Scenario 2(a), that is, the performance of the SVCI model with the logistic regression likelihood or with a larger $m$ is more preferable. Table~\ref{tab4} displays the results from LGCP as a comparison, which indicate a clear improvement of using SVCI over LGCP in terms of estimation accuracy.

\begin{table}[]
    \centering
    \caption{Scenario 2: mean integrated squared error of log intensity  ($\text{MISE}_{\log\rho}$) averaged over $100$ simulations for different values of $m$. We compare the two estimating equations, the Poisson likelihood (PL) and the logistic regression likelihood (LRL) with their competitors.}
    \begin{tabular}{cccccc}
        \hline\hline
        \multirow{2}{*}{Method}&\multicolumn{5}{c}{$\text{MISE}_{\log\rho}$}\\
        \cline{2-6}
        &{$m=800$}&{$m=1600$}&{$m=2400$}&{$m=3600$}&{$m=6000$}\\
        \hline\hline
        \multicolumn{6}{c}{(a): \quad $\rho(u)=\exp\{\beta_0(u)\}$}\\
        \hline
        SVCI-PL&0.128&0.101&0.084&0.057&0.041\\
        SVCI-LRL&0.117&0.095&0.074&0.042&0.030\\
        KDE.lpp&0.157&0.140&0.112&0.084&0.061\\
        KDEQuick.lpp&0.133&0.127&0.109&0.075&0.054\\
        Voronoi.lpp&0.128&0.120&0.094&0.067&0.048\\
        \hline
        \multicolumn{6}{c}{(b): \quad  $\rho(u)=\exp\{z_1(u)\beta_1(u)+z_2(u)\beta_2(u)+\beta_0(u)\}$}\\
        \hline
        SVCI-PL&0.177&0.152&0.135&0.114&0.085\\
        SVCI-LRL&0.165&0.141&0.122&0.097&0.072\\
        LGCP&0.227&0.202&0.188&0.154&0.137\\
        \hline
    \end{tabular}
    \label{tab4}
\end{table}


\section{Real Data Analysis}\label{sec:realdata}
 
 We consider two real data examples to illustrate the performance of the proposed method. The first Toronto Homicide data example has a moderate data size with $1398$ points and three explanatory variables on a domain with irregular boundaries. And the second Western Australia Traffic Accidents data has a larger data size with $14,562$ points on a linear road network. In both studies, we use SVCI-LRL and $\texttt{nd}^2\approx m$, due to their favorable performance in our simulation studies. 

\subsection{Toronto Homicide Dataset}
We apply the proposed SVCI model to the analysis of the Toronto Homicide dataset. The raw dataset contains the information of $1398$ homicides occurred in Toronto, Canada during 1990 to 2014, recording the locations of murder scenes, homocide types and information of victims obtained from the Toronto Star Newspaper (\url{http://www.thestar.com/news/crime/torontohomicidemap.html}). We select the more recent years since 2000 and delete the data which have duplicated locations. There remains $764$ homicide cases for the final analysis. Figure \ref{fig1} shows the entire Toronto city and the locations of the selected cases within a $42\times 31$ km rectangle window.  
Notably, the old Toronto region in the middle of the coast has more frequent occurrences of homicides. 

\begin{figure}
\begin{center}
\includegraphics[width=6in]{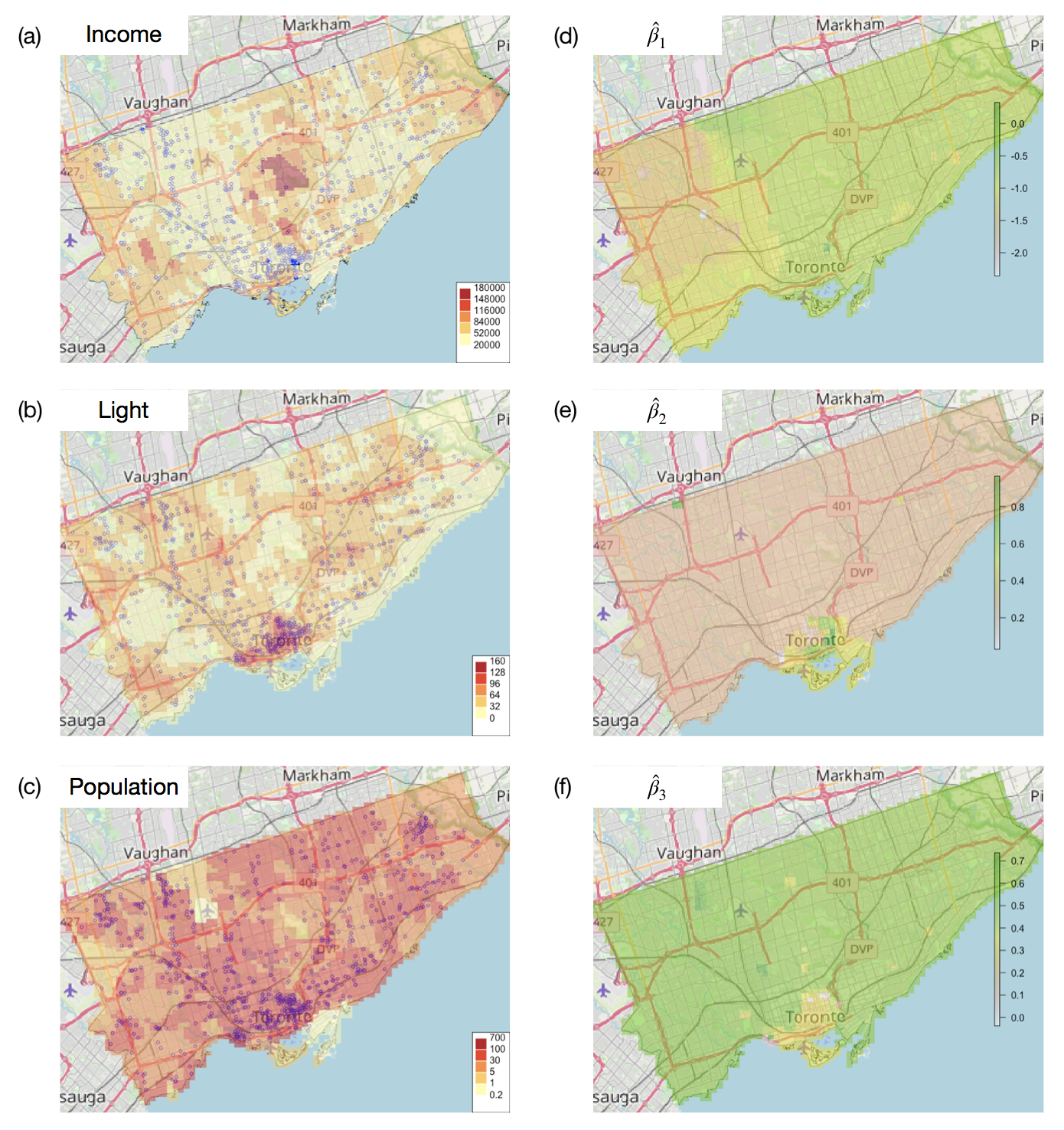}
\end{center}
\caption{Left Panel: patterns of covariates; Right Panel: patterns of estimates; (a,d) Average income of the residents in Toronto; (b,e) Light intensity in Toronto night; (c,f) Population density in Toronto;}
\label{fig3}
\end{figure}

The data also contains the records of average income, night lights and population density of Toronto city in 2006, and we use them as the explanatory variables. Figures \ref{fig3} (a-c) show the observations of the three variables. As can be seen, there is a large spatial variation of these variables across the city. We scaled and centered each spatial covariates before running our point process models. 

We focus on the investigation of the relationship between the distribution of homicides and the three explanatory variables. We first fit a standard log-Gaussian cox process model with constant regression coefficients as a benchmark for comparisons. The intensity function of LGCP takes the form
\begin{equation*}
    \log\{\rho(u)\}=\beta_0+\text{Income}(u)\beta_1+\text{Night}(u)\beta_2+\text{Pdens}(u)\beta_3+\phi(u)
\end{equation*}
where $\beta_1$, $\beta_2$ and $\beta_3$ are the constant-coefficients across the domain and $\phi(u)$ is a spatial Gaussian process with a zero mean and a Matern correlation function. After centering and scaling the covariates, we obtain the parameter estimates from LGCP as $\beta_1=-0.918$, $\beta_2=0.378$, $\beta_3=0.207$ and $\beta_0=1.096$. These estimates imply that the homicides are more likely to occur in the area with a lower average income, a better lights condition and a denser residential population. 

We then fit the SVCI model with a $5$-NN graph assuming that the homicide locations follow a spatial point process with the following intensity function,
\begin{equation*}
    \log\{\rho(u)\}=\text{Income}(u)\beta_1(u)+\text{Night}(u)\beta_2(u)+\text{Pdens}(u)\beta_3(u)+\beta_0(u)
\end{equation*}
Here $\beta_k(u)$, $k=0,1,2,3$ are spatially varying coefficients, whose estimates are shown in Figure \ref{fig3}(d-f). It takes about $0.072$ seconds to construct the $5$-NN graph and $0.69$ seconds to get an estimate of $\bbeta$ for each tuning parameter. 
Clearly, the results of SVCI reveal more details about the effects of covariates than those from LGCP.
Overall, the signs of $\beta_k(u)$ are the same as the results of LGCP. For example, the estimates of $\bbeta_1(u)$ range from -2 to 0, indicating a negative relationship between income and homicide occurrence as is expected. Such a negative relationship is most prominent in the western region of Toronto whereas a weaker relationship is observed near the upper east corner of Toronto City. 
For both $\beta_2$ and $\beta_3$, we observe a small cluster at the Old Toronto region, which has the most concentrated homicide cases. It is notable that the relationships between light intensity/population density and homicides occurrence in the Old Toronto region differ significantly from the rest of Toronto city; a stronger positive relationship is observed for both variables.

\subsection{Western Australia Traffic Accidents}
In this section, we study the traffic accidents data in the state of Western Australia for the year 2011, as shown in Figure \ref{fig1}. The data were originally provided by the Western Australian State Government Department of Main Roads and are made publicly available as part of the Western Australian Whole of Government Open Data Policy. The data can also be accessed from the \texttt{R} package \texttt{spatstat.Knet}.
It consists of $14,562$ locations of accidents on a road network with $115,169$ road segments constrained in a $[217.4, 1679.1]\times[6114.9, 7320.6]$ km rectangle window.  

The grey lines in Figure \ref{fig1} represent the traffic network of Western Australia and each red point denotes an accident spot. It is clear from this Figure that accidents are highly concentrated around the Perth metropolitan area located in the western coastal region. This region contains nearly $75\%$ of the population in Western Australia. By contrast, the remote eastern region has a much sparser road network and a smaller number of traffic accidents. Our goal is to estimate the intensity function over this network to investigate the spatial variation of accident occurrences.   
  
\begin{figure}
\begin{center}
\includegraphics[width=4in]{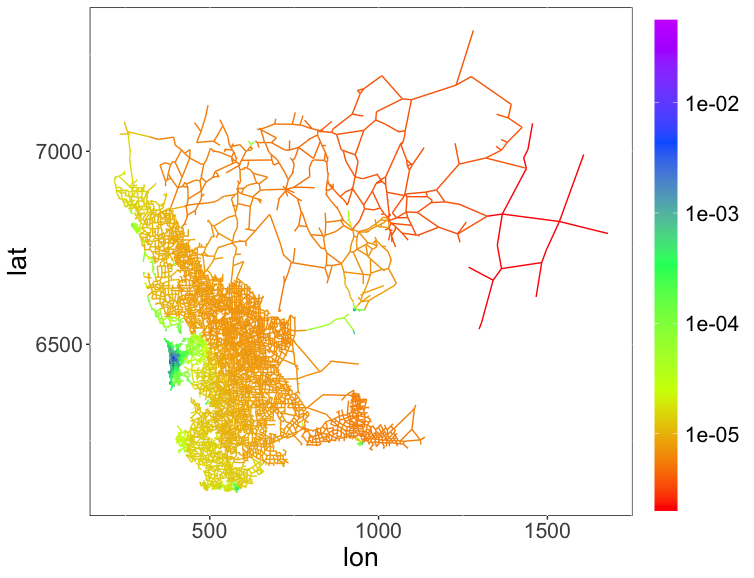}
\end{center}
\caption{Intensity estimates for the accidents on the Western Australian road network.}
\label{fig4}
\end{figure}

We build the SVCI  model of the intensity function with a spatially varying intercept, $\rho(u;\beta)=\exp\{\beta_0(u)\}$. In this study, we don't have any spatial covariates available and hence we focus on detecting the clustered patterns of the intensity function $\rho(u)$. 
SVCI takes about $1.75$ minutes to construct a connection graph using the graph construction method in Section 3.1 and takes on average $5.01$ seconds to get an estimate of $\bbeta$ for each tuning parameter. 
Figure \ref{fig4} plots the estimated log intensity $\log\hat{\rho}(u)$ on the road network. We notice that $\hat{\rho}(u)$ has a large spatial variation, ranging from $0$ per kilometer in some remote eastern areas to nearly $50$ accidents per kilometer in some busy roads in the Perth metropolitan area. 

\begin{figure}
\begin{center}
\includegraphics[width=6.3in]{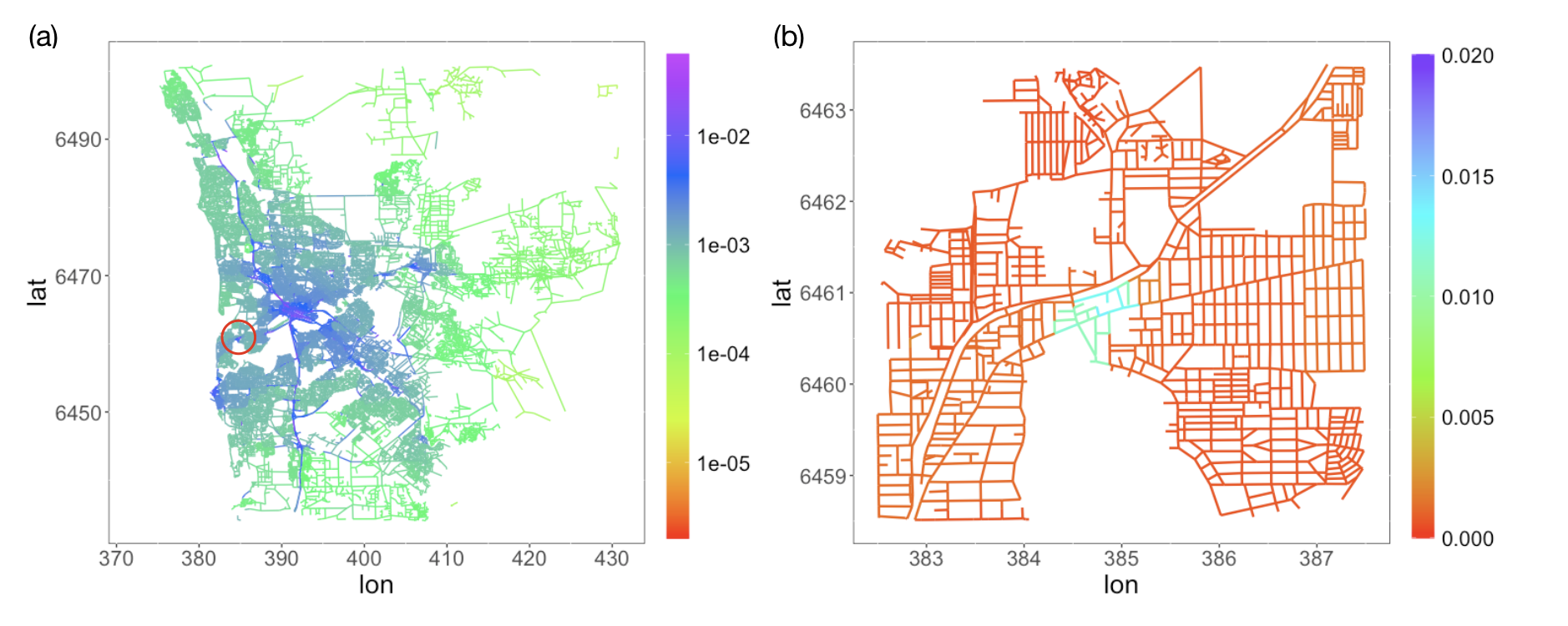}
\end{center}
\caption{(a): Intensity estimates map for the accidents in the metropolitan Perth Area; (b): Intensity estimates map by zooming into the red circle area in the left panel.}
\label{fig5}
\end{figure}

We zoom into the sub-region of $[372, 431]\times [6434, 6501]$ km, which is displayed in Figure \ref{fig5} (a), to have a detailed investigation of the traffic accident rates in the densely populated Perth metropolitan area.  It is clear that several roads are having substantially higher intensities than the rest of the roads, many of which are along the major freeways of the city.  In particular, we observe very high intensity values at or near the center of the city marked by the purple color. Indeed, these roads and intersection are located at the Perth Central Business District.  In contrast, although having dense local road networks, many residential areas away from highways have relatively lower intensity values. One advantage of SVCI lies in its capability of capturing intensity functions with abrupt changes. To give an example, we highlight a road segment on Highway 5 in Perth by a red circle in Figure \ref{fig5} (a), and show the zoomed map in Figure~\ref{fig5} (b). It is noticeable that a sudden jump in the estimated intensity function appears near the southwest end of the road. After verifying with the Google satellite image, we confirm that the northeast part of the road passes through a large residential area, whereas the southwest part is a commercial and public service area (restaurants/shops/school/church/hospital) that is expected to have a higher rate of accidents.

\section{Conclusion}\label{sec:conclusion}
In this study, we propose a varying coefficient log-linear intensity model, referred to as the SVCI model,  for the visualization and analysis of spatial point processes. We utilize a graph fused lasso regularization to estimate the clustering patterns of the regression coefficients. The method guarantees spatially contiguous clustering configurations with highly flexible cluster shapes and data-driven cluster sizes. It supplements the current research on intensity estimation, which primarily focuses on relatively smooth intensity functions without covariates or spatially constant regression coefficients. The method also has the advantage of being applicable to a broad range of complex domains such as line networks and spatial domains with irregular boundaries. The computation of the model is made highly efficient by using a proximal gradient optimization algorithm. Numerical studies show that our method produces more accurate intensity estimations than several competing methods such as the KDE-based methods \citep{mcswiggan2017kernel,rakshit2019fast} and the resample-smoothing Voronoi intensity estimation method \citep{moradi2019resample}, when intensity functions exhibit discontinuous changes on linear networks. The computation of SVCI is in general
reasonable compared to its competitors considered in this paper. The method is applied to identify spatially heterogeneous patterns in the determinants of Toronto crime events and the intensity of traffic accidents in Western Australia.

Moving forward, this work could be further refined in several ways. First, SVCI only considers a small fixed number of covariates. However, in practice, practitioners may face a large number of available covariates but lack a strong theory to inform variable selection. There is a need of a more general model that allows researchers to undergo variable selection and spatial cluster identification simultaneously for point processes. Second, the SVCI estimator does not come with an uncertainty measure that makes it hard for statistical inference, a common issue shared by many regularization based approaches. We may consider a Bayesian version of the method or a bootstrapping based approach to address the inference problem. Third, an interesting research direction is to extend the finite dimensional graph regularization based method to an infinite dimensional process-based clustered coefficient model such that spatial predictions can be done in a more rigorous way. Moreover, the method can be extended by considering a weighted composite log-likelihood to further improve statistical efficiency for non-Poisson point processes \citep{guan2010weighted}. Finally, we acknowledge that further investigations are needed on how to verify and relax some of the assumptions that are used to establish the theoretical results. Tighter error bounds may be obtained following the method of entropy bound similar as those in \cite{ortelli2019oracle}. In addition, infilling domain asymptotic theoretical framework might be more reasonable for point patterns on bounded domains or linear networks. However, these are not trivial theoretical questions, and we will leave those for future research. 

\setcounter{section}{0}
\setcounter{figure}{0}

\counterwithin*{equation}{section}
\counterwithin*{equation}{subsection}
\counterwithin*{table}{section}
\counterwithin*{table}{subsection}

\renewcommand{\theequation}{A\arabic{equation}}
\renewcommand{\thetable}{A\arabic{table}}
\renewcommand{\thefigure}{A\arabic{figure}}
\renewcommand\thesection{A\arabic{section}}
\renewcommand\thesubsection{A\arabic{section}.\arabic{subsection}}

\section*{Appendix}

\input{SpatStat_V1/parts/Appendix_Theory}

\bibliographystyle{elsarticle-harv}
\biboptions{authoryear}
\bibliography{Ref}
\end{document}

%% file: SpatStat_V1/parts/Appendix_Theory.tex
\section{Proof of Theorem 1}\label{Appendix1}
We state two useful lemmas to be used in the proof of Theorem~\ref{eq:theorem}. 
\begin{lemma}\label{Lemma:1}
Assume that Assumptions $1-4$ hold, 
then

$$
\left|D_{n}\right|^{1 / 2} U_{n}^{(1)}\left(\bbeta^{0}\right) \stackrel{p}{\rightarrow} N\left(\mathbf{0}, \bfB_1+\bfC_1\right)
$$ 
\end{lemma}
The proof of Lemma~\ref{Lemma:1} can be found in \cite{guan2010weighted} and \cite{thurman2015regularized}.

\begin{lemma} \label{Lemma:2}
If $\epsilon_1,\dots, \epsilon_n$ are sub-Gaussian random variables that $P(\epsilon_i \geq t) \leq \exp(-\frac{t^2}{2\sigma_i^2})$ . $\epsilon_1,\dots, \epsilon_n$ do not need to be independent. We denote $\sigma^2 = \max_i(\sigma^2_i)$ and there exists $C$ that
\begin{equation*}
    P(\max_{i} |\epsilon_i| \geq  t) \leq C n\exp(-\frac{t^2}{2\sigma^2}).
\end{equation*}
\end{lemma}
Lemma 2 is a standard result for sub-Gaussian variables which can be easily derived using the union bound.

\begin{proof}[\textbf{Proof of Theorem 1}]
Note that the negative Poisson based composite log-likelihood function takes the form $$U_n(\bbeta)=-\{\frac{1}{|D_n|} \sum_{\mathbf{u} \in D_n \cap \bfX} \log \rho(u ; \bbeta)-\frac{1}{|D_n|} \int_{D_n} \rho(u ; \bbeta) d u\}.$$ 
Then
\begin{eqnarray*}
U_{n}^{(1)}\left(\bbeta\right)
&= &  -\frac{1}{|D_n|}(\sum_{u \in D_n \cap \bfX} z(u)-\int_{D}  z(u) \rho\left(u ; \bbeta\right) d u) \\ 
 U_{n}^{(2)}(\bbeta)&=& \frac{1}{|D_n|}\int_{D_n}  z(u) z(u)^{T} \rho(u ; \bbeta) d u =|D_n|^{-1}\bfB_{n}(\bbeta)
\end{eqnarray*}
 
From Assumption 5, it follows that  there exists a constant $m^{\prime}$ such that
\begin{eqnarray}\label{eq:convex}
U_{n}(\bbeta)-U_{n}\left(\bbeta^{0}\right)- U_n^{(1)}\left(\bbeta^{0}\right)^{\top}\left(\bbeta-\bbeta^{0}\right) \geq \frac{m^{\prime}}{2}\left\|\bbeta-\bbeta^{0}\right\|_{2}^{2}, \quad \text { for } \bbeta \in N_{\delta}(\bbeta^0)
\end{eqnarray}
Recall that $\widehat{\bbeta}$ minimizes the penalized negative log likelihood function $Q(\bbeta)$. 
We have the basic inequality  
$$
U_n(\widehat{\bbeta})+\lambda\|H \widehat{\bbeta}\|_{1} \leq U_n\left(\bbeta^{0}\right)+\lambda\left\|H \bbeta^{0}\right\|_{1}
$$

Combine the above basic inequality with the inequality in \eqref{eq:convex},  we have
\begin{eqnarray}
\label{eq:inequal2}
\frac{m^{\prime}}{2}\left\|\widehat{\bbeta}-\bbeta^{0}\right\|_{2}^{2} \leq -U_{n}^{(1)}\left(\bbeta^{0}\right)^{\top}\left(\widehat{\bbeta}-\bbeta^{0}\right)+\lambda\left(\left\|H \bbeta^{0}\right\|_{1}-\|H \widehat{\bbeta}\|_{1}\right)
\end{eqnarray}

We rewrite the inequality in \eqref{eq:inequal2} as 
$$
\frac{m^{\prime}}{2}\left\|\widehat{\bbeta}-\bbeta^{0}\right\|_{2}^{2} \leq -U_{n}^{(1)}\left(\bbeta^{0}\right)^{\top}P_{{H}^{\perp}}H^{\dagger}H\left(\widehat{\bbeta}-\bbeta^{0}\right)+\lambda\left(\left\|H \bbeta^{0}\right\|_{1}-\|H \widehat{\bbeta}\|_{1}\right)
$$

Applying the Holder's inequality to the first term on the right side, we obtain
$$
\left\|-U_{n}^{(1)}\left(\bbeta^{0}\right)^{\top}P_{{H}^{\perp}}H^{\dagger}H\left(\widehat{\bbeta}-\bbeta^{0}\right)\right\|_1 \leq \left\|U_{n}^{(1)}(\bbeta^{0})^{T}P_{{H}^{\perp}}H^{\dagger}\right\|_{\infty}\left\|H\left(\tilde{\bbeta}-\bbeta^{0}\right)\right\|_{1}
$$

Let $\lambda \geq \bar{\lambda}= \|\{P_{{H}^{\perp}}H^{\dagger}\}^{T}U_{n}^{(1)}(\bbeta^{0})\|_{\infty}$. Using the triangular inequality, we obtain that
\begin{equation}
\frac{m^{\prime}}{2}\left\|\widehat{\bbeta}-\bbeta^{0}\right\|_{2}^{2} \leq 4 \lambda\left\|H \bbeta^{0}\right\|_{1}
\end{equation}

Below we will derive the rate for $\bar{\lambda}=\|\{P_{{H}^{\perp}}H^{\dagger}\}^{T}U_{n}^{(1)}(\bbeta^{0})\|_{\infty}.$
From Lemma 1, we have 
\begin{equation}\label{eq:inequal3}
\left|D_{n}\right|^{1 / 2} U_{n}^{(1)}\left(\bbeta^{0}\right) \stackrel{p}{\rightarrow} N\left(0, \bfB_1+\bfC_1\right) 
\end{equation}

Let $\boldsymbol{\varepsilon}=|D_{n}|^{1 / 2}(\bfB_1+\bfC_1)^{-1/2}U_{n}^{(1)}\left(\bbeta^{0}\right)$. It follows that $\boldsymbol{\varepsilon}$ is a standard normal vector.   
We rewrite $\bar{\lambda}$ as $$\|\{P_{{H}^{\perp}}H^{\dagger}\}^{T}U_{n}^{(1)}(\bbeta^{0})\|_{\infty}=|D_n|^{-1/2}\|\{(\bfB_1+\bfC_1)^{1/2}P_{{H}^{\perp}}H^{\dagger}\}^{T} \boldsymbol{\varepsilon}\|_{\infty}$$
Let $\mathbf{v}_{k}$ be the $k$th column of $(\bfB_1+\bfC_1)^{1/2}P_{{H}^{\perp}}H^{\dagger}$. From Lemma 2, we can then prove that $\bar{\lambda}=O(|D_n|^{-1/2}M^{*}\sqrt{\log Mp})$, where $M^{*}=\sqrt{\max_{k}\left\|\mathbf{v}_{k}\right\|^{2}}$. 
\end{proof}


\section{Quadrat Scheme}\label{Appendix2}
\begin{figure}
\begin{center}
\includegraphics[width=5in]{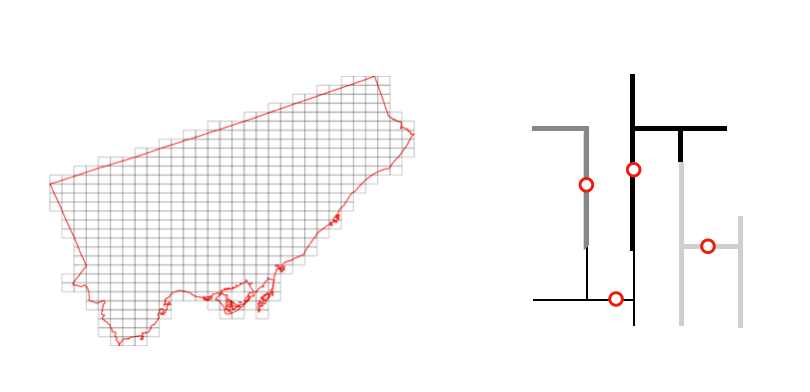}
\end{center}
\caption{Illustration of quadrat schemes for Toronto city (left) and a toy linear network example (right); Left panel: boundary lines (red) of Toronto city and the grids (grey) covering the irregular domain; Right panel: dummy points (red nodes) and their corresponding Voronoi quadrats (marked by different shades of grey).}
\label{app_fig1}
\end{figure}

For the quadrature approximation in (\ref{app1}), we need to divide the domain into small subdivisions or quadrats. For the 2-D square domain, it is easy to add dummy points and draw regular equal-sized rectangular quadrats on the domain using $\texttt{dummy.ppm}$ and $\texttt{quad.ppm}$ in $\texttt{R}$ library $\texttt{spatstat}$. 
For planar domains with irregular boundary, we follow the routine quadrature approximation methods for irregular domains in $\mathbb{R}^2$ \citep{shen2009local}, where the domain is masked by 
regular pixel grids, and we expand the domain slightly to include the whole pixel if it has intersections with domain boundary. See the left panel of Figure \ref{app_fig1} for an example.  

As for the partition on a linear work, Chapter 9 of \cite{okabe2012spatial} discussed the implementation of both equal-length and unequal-length network cells. \cite{furuta2008voronoi,shiode2008analysis} proposed computational methods for dividing a network into equal-length network cells. However, their methods are not a guarantee of success if we want to insert enough dummy points. 
We propose to randomly draw dummy points from a homogeneous Poisson point process on the linear network with intensity function $\delta(u)=(M-m)/|D|$. Then we obtain the quadrats on linear networks using the network Voronoi tessellation method \citep[Chapter 4 of] []{okabe2012spatial}, each of which contains one dummy point as its centroid. See the right panel of Figure \ref{app_fig1} for an example.  

 


\section{Addition Numerical Results}\label{appsec:sim}
We report the integrated squared bias (ISB) and variance (IV) for each $\hat{\bbeta}_1$, $\hat{\bbeta}_2$ and $\hat{\bbeta}_0$ under both Scenario 1 and Scenario 2(b) in Table \ref{app_tab1}. Specifically, ISE and IV are defined as
 $$\text{ISB}(\bbeta_i)=\frac{1}{|D|}\int_D(\hat{\beta}(u)-\mathbb{E}\hat{\beta}(u))^2,\qquad \text{IV}(\bbeta_i)=\text{MISE}(\bbeta_i)-\text{ISB}(\bbeta_i).$$
The results generally agree well with the findings based on Rand index and the $\text{MISE}_{\beta}$, in the sense that SVCI-LRL achieves a slightly smaller bias and variance compared with SVCI-PL. 

\begin{table}[!ht]
    \centering
    \caption{Scenario 1: The integrated squared bias and variance (in parentheses) of $\hat{\bbeta}_1$, $\hat{\bbeta}_2$ and $\hat{\bbeta}_0$ respectively, in the case that $m=2400$, and $\texttt{nd}^2=m$ based on $5$-NN connection graphs.}
    \begin{tabular}{cccc}
         \hline\hline
         &$\bbeta_1$&$\bbeta_2$&$\bbeta_0$\\
         \hline\hline
         \multicolumn{4}{c}{Scenario 1}\\
         \hline
         SVCI-PL&0.115(0.035)&0.116(0.032)&0.078(0.038)\\
         SVCI-LRL&0.098(0.034)&0.105(0.033)&0.057(0.030)\\
         \hline
         \multicolumn{4}{c}{Scenario 2(b)}\\
         \hline
         SVCI-PL&0.099(0.028)&0.104(0.030)&0.108(0.035)\\
         SVCI-LRL&0.105(0.028)&0.101(0.027)&0.093(0.030)\\
         \hline
    \end{tabular}
    \label{app_tab1}
\end{table}

\begin{figure}
\begin{center}
\includegraphics[width=3.5in]{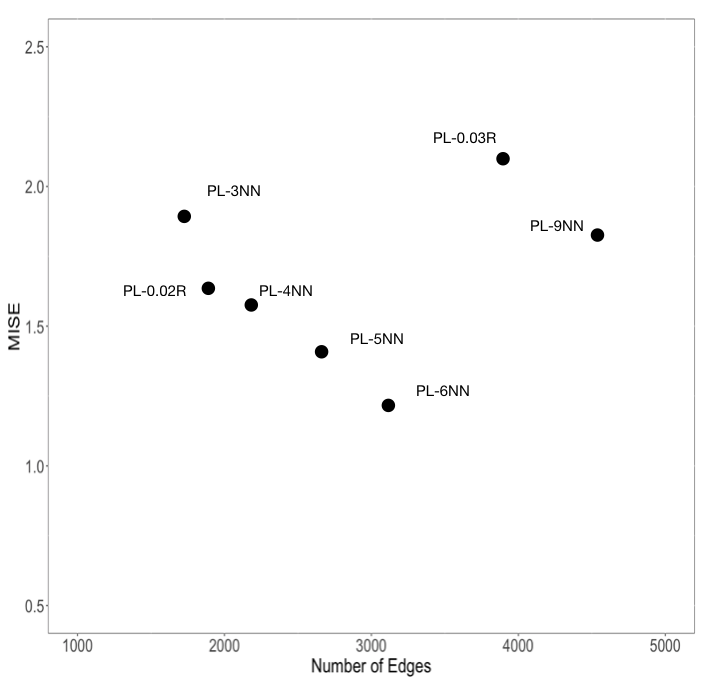}
\end{center}
\caption{Illustration of $\text{MISE}_{\beta}$ versus the number of edges, under the setting of Scenario 1, with $n=2400$ and $\texttt{nd}^2=m$. 
$\epsilon$R indicates the SVCI-PL method based on an $r$-NN graph with a radius $\epsilon\times R$.}
\label{app_sim1}
\end{figure}

Under the setting of Scenario 1, we examine the performance of SVCI-PL based on $K$-NN graphs with $K=3,4,5,6,9$ and $r$-NN graphs with $r=0.02R, 0.03R$. Figure \ref{app_sim1} shows the $\text{MISE}_{\beta}$ versus the number of edges for each graph. We notice that MISE does not always decrease as the graph includes more neighbors by increasing $K$ of $K$-NN or $r$ of $r$-NN. 
Both $K$-NN and $r$-NN lose some estimation accuracy if there are too many neighbors. It also seems that $K$-NN slightly outperform $r$-NN in terms of MISE when $K$ and $r$ are chosen such that the two graphs have comparable number of edges. 

Under the setting of Scenario 2(a), we compare the performance of SVCI-LRL using the $3$-NN graphs constructed based on the shortest-path distance and Euclidean distance metrics, respectively. Overall, we observe very similar results between these two choices of distance metrics, especially when $m$ and $\texttt{nd}^2$ go up. For example, when $n=2400$, the $\text{MISE}$s of SVCI-LRL using shortest-path distance and Euclidean distance metrics are $0.121$ and $0.124$, respectively, which are very close to the result reported in Table \ref{tab4} obtained using the graph constructed by connecting natural neighbors on the Chicago network.

Table \ref{app_tab2} reports the computation time of different methods with one tuning parameter, under the setting of Scenario 2(a). We notice that the computation time of KDE.lpp and Voronoi.lpp vary notably with their tuning parameter, so the computation times are calculated by averaging $20$ repeats with a sequence of tuning parameters. 

\begin{table}[]
    \centering
    \label{app_tab2}
    \caption{Scenario 2: Comparison of computation times (in seconds).}
    \begin{tabular}{cccccc}
        \hline\hline
        Method &{$m=800$}&{$m=1600$}&{$m=2400$}&{$m=3600$}&{$m=6000$}\\
        \hline\hline
        \multicolumn{6}{c}{Scenario 2(a): \quad $\rho(u)=\exp\{\beta_0(u)\}$}\\
        \hline
        SVCI-LRL&$0.64$&$0.73$&$1.11$&$1.40$&$1.92$\\
        KDE.lpp&$6.87$&$8.26$&$10.96$&$13.16$&$16.17$\\
        KDEQuick.lpp&$0.057$&$0.066$&$0.084$&$0.092$&$0.102$\\
        Voronoi.lpp&$3.61$&$4.02$&$4.95$&$5.48$&6.83\\
        \hline
    \end{tabular} 
\end{table}

\begin{figure}
\begin{center}
\includegraphics[width=6in]{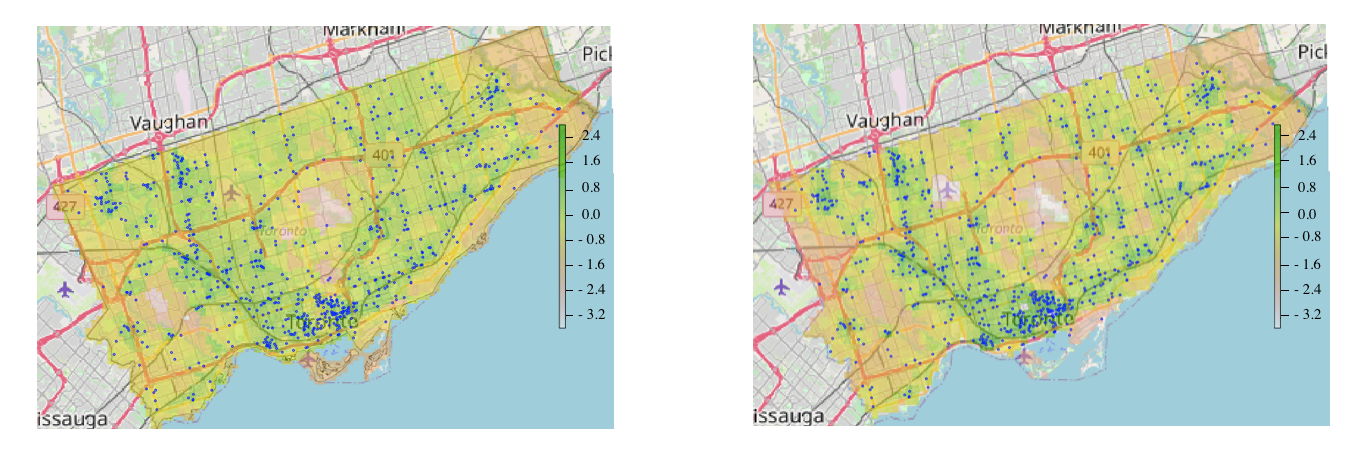}
\end{center}
\caption{Illustrate of the estimated log intensity function of the Toronto data with the observed locations overlaid in blue dots; Left: the estimated log intensity surface by SVCI; Right: the posterior mean estimate of the log intensity surface by LGCP;}
\label{app_real11}
\end{figure}

For the Toronto Homicide data considered in Section 5.1, we compare the estimated log intensity surfaces obtained by SVCI and LGCP in Figure \ref{app_real11}. Both methods seem to be capable of capturing the inhomogeneity pattern in intensities and agree well with each other in most areas. The most notable difference between the two methods occurs near Toronto islands, where we observe more variations in intensity estimations by LGCP than those by SVCI possibly due to the piece-wise homogeneity assumptions made in the latter. 



